\documentclass[a4paper,twocolumn,english,showpacs,nofootinbib,nobibnotes,aps,prl,floatfix,superscriptaddress]{revtex4-2}
\usepackage[utf8]{inputenc}
\usepackage{amsmath}
\usepackage{amsthm}
\usepackage{amssymb}
\usepackage{graphicx}
\usepackage[caption=false]{subfig}
\usepackage{xparse}
\usepackage{xcolor}
\usepackage{hyperref}

\makeatletter

\usepackage{braket}
\usepackage{dsfont}

\usepackage{notes2bib}

\bibnotesetup{
  note-name = ,
  use-sort-key = false
}

\makeatother

\usepackage{babel}

\newcommand{\trace}{\operatorname{Tr}}

\newcommand{\one}{\mathds{1}}

\newcommand{\RR}[2]{\mathcal{R}_{#2}^{(#1)}}

\NewDocumentCommand\opti{smmm>{\SplitList{;}}m} {
\begingroup%
\setlength{\belowdisplayskip}{-0.6\baselineskip}%
\IfBooleanTF{#1}{%
    \begin{alignat*}{2}
        & \underset{#3}{\text{#2}} & & #4 \\
        & \text{subject to~~}
        \ProcessList{#5}{ \insertopticonst }
        & &
    \end{alignat*}%
    }{%
    \begin{alignat}{2}
        & \underset{#3}{\text{#2}} & & #4 \\
        & \text{subject to~~}
        \ProcessList{#5}{ \insertopticonst }
        & & \nonumber
    \end{alignat}%
    }%
\endgroup%
}%
\newcommand\insertopticonst[1]{& & #1\\&}

\begin{document}

\title{Complete characterization of quantum correlations by randomized measurements}
\author{Nikolai Wyderka}
\affiliation{Institut für Theoretische Physik III, Heinrich-Heine-Universität Düsseldorf, Universitätsstr.~1, 40225 Düsseldorf, Germany}
\author{Andreas Ketterer}
\affiliation{Fraunhofer Institute for Applied Solid State Physics IAF, Tullastr.~72, 79108 Freiburg, Germany}
\author{Satoya Imai}
\affiliation{Naturwissenschaftlich-Technische Fakultät, Universität Siegen, Walter-Flex-Str.~3, 57068 Siegen, Germany}
\author{Jan Lennart Bönsel}
\affiliation{Naturwissenschaftlich-Technische Fakultät, Universität Siegen, Walter-Flex-Str.~3, 57068 Siegen, Germany}
\author{Daniel E. Jones}
\affiliation{DEVCOM Army Research Laboratory, Adelphi, Maryland 20783, USA}
\author{Brian T. Kirby}
\affiliation{DEVCOM Army Research Laboratory, Adelphi, Maryland 20783, USA}
\affiliation{Tulane University, New Orleans, Louisiana 70118, USA}
\author{Xiao-Dong Yu}
\affiliation{Naturwissenschaftlich-Technische Fakultät, Universität Siegen, Walter-Flex-Str.~3, 57068 Siegen, Germany}
\author{Otfried Gühne}
\affiliation{Naturwissenschaftlich-Technische Fakultät, Universität Siegen, Walter-Flex-Str.~3, 57068 Siegen, Germany}
\date{\today}

\begin{abstract}
The fact that quantum mechanics predicts stronger correlations than classical physics is an essential cornerstone of quantum information processing. Indeed, these quantum correlations are a valuable resource for various tasks, such as quantum key distribution or quantum teleportation, but characterizing these correlations in an experimental setting is a formidable task, especially in scenarios where no shared reference frames are available. By definition, quantum correlations are reference-frame independent, i.e., invariant under local transformations; this physically motivated invariance implies, however, a dedicated mathematical structure and, therefore, constitutes a roadblock for an efficient analysis of these correlations in experiments. Here we provide a method to directly measure any locally invariant property of quantum states using locally randomized measurements, and we present a detailed toolbox to analyze these correlations for two quantum bits. We implement these methods experimentally using pairs of entangled photons, characterizing their usefulness for quantum teleportation and their potential to display quantum nonlocality in its simplest form. Our results can be applied to various quantum computing platforms, allowing simple analysis of correlations between arbitrary distant qubits in the architecture.

\end{abstract}

\maketitle

{\it Introduction.---}
Quantum mechanics contains a plethora of fascinating nonlocal effects that are useful in various applications of quantum technologies. Such effects are, by definition, 
invariant under changes of the local reference systems or, mathematically speaking, of the choice of the local bases of the Hilbert space. This naturally leads to the expectation that they should be describable by quantities which are invariant under such transformations. 
A quantum state of composite systems is described by a density matrice $\rho_{AB}$ in the tensor product space of the individual systems, so invariance under local basis changes of any function $f(\rho_{AB})$ of the state can, in the case of bipartite systems, be expressed as
\begin{align}
    f(\rho_{AB}) = f(U_A\otimes U_B \rho_{AB} U_A^\dagger \otimes U_B^\dagger),
\end{align}
where $U_A$ and $U_B$ are unitary matrices governing the basis change of the first and second system, respectively. Due to this invariance, an average over all such transformations yields
\begin{align}
    f(\rho_{AB}) = \iint \text{d}U_A\text{d}U_B f(U_A\otimes U_B \rho_{AB} U_A^\dagger \otimes U_B^\dagger).
\end{align}
On the other hand, any physical function may be expanded in terms of powers of expectation values of certain 
observables, yielding 
\begin{align}
f(\rho_{AB}) = \sum_{\vec{t}} c_{\vec{t}}\, \langle \mathcal{M}_1 \rangle^{t_1} \langle \mathcal{M}_2 \rangle^{t_2} \ldots
\end{align}
for appropriately chosen bipartite observables $\mathcal{M}_i$ and coefficients $c_{\vec{t}}$, where $\vec{t} = (t_1, t_2, \dots)$ denotes all possible multi-indices with positive integers $t_i$ and a varying
number of entries $k$. Combining this with local unitary invariance yields
\begin{align}
\label{eq:moments}
f(\rho_{AB}) = \sum_{\vec{t}} c_{\vec{t}}\, \RR{t_1}{\mathcal{M}_1}(\rho_{AB})\RR{t_2}{\mathcal{M}_2}(\rho_{AB})\ldots,
\end{align}
where the quantities
\begin{align}\label{eq:mom}
    \RR{t}{\mathcal{M}}(\rho) := \!\!\iint \text{d}U_A \text{d}U_B \{\trace[(U_A\otimes U_B) \rho (U_A^\dagger \otimes U_B^\dagger) \mathcal{M}]\}^t.
\end{align}
are the $t$-th moments of the probability distribution of measurement results for the bipartite observable $\mathcal{M}$ under random local basis changes. For the case of product observables, the quantities $\RR{t}{A\otimes B}$ have been studied as 
randomized measurements, 
see, e.g., Refs.~\cite{PhysRevA.92.050301,Knips:2020aa,Knips2020momentrandom,elben2022randomized, cieslinski2023analysing}.
The advantages of these schemes include the possibility to obtain the data without having shared reference frames, having limited control over the measurements and in the presence of uncharacterized local unitary noise.  With sufficient experimental control, the random unitaries may even be selected from a finite unitary $t$-design instead \cite{gross2007evenly,Ketterer2020entanglement}. Note that the mild assumptions on the measurement
capabilities are in contrast to the stronger ones in shadow tomography schemes, where known random unitary rotations are applied to a state to estimate expectation values with a small number of measurements \cite{huang2020predicting}.

Here, we go beyond the standard randomized measurement schemes by allowing for non-product observables. While this sounds like a disadvantage for practical implementations, we stress that it is possible to obtain the moment data for non-product observables from the data of multiple product observables by classical post-processing.

As the moments of these distributions can be measured directly, they form the main objects of interest in order to describe the local unitary invariant functions. In principle, it is possible to expand any (polynomial) local unitary invariant (LU invariant) in this manner. However, so far only a small subset of these moments 
has been exploited for tasks like entanglement detection \cite{brydges2019probing, elben2020mixed,Ketterer_2019,imai2021bound} or fidelity estimation~\cite{PhysRevLett.106.230501,Elben_2020}. 

In this paper, we develop a general framework linking the moments of randomized measurements and the set of LU invariants. For fixed local dimension $d$, the set of polynomial invariants is finitely generated
\cite{grassl1998computing, springer2006invariant}, so it suffices to consider the generators. Indeed, complete sets of generators have been found in the case of two-qubit states and certain classes of higher-dimensional cases \cite{makhlin2002nonlocal, sun2017local}. In the following, we develop concrete schemes to measure {\it all} of the relevant two-qubit invariants, but naturally our theory can be extended to known invariants in higher-dimensional or multiparticle systems. We illustrate this by deriving a scheme to measure the Kempe invariant in three-qubit systems \cite{kempe1999multiparticle, barnum2001monotones}.
As an application, we experimentally implement a randomized measurement scheme to measure some of the invariants and use it to certify the presence of Bell nonlocality and the usefulness of the prepared states for teleportation schemes.

\vspace{1em}
{\it Randomized measurements.---}
In the framework of randomized measurements, a multiparticle quantum state undergoes random local unitary transformations before a fixed observable $\mathcal{M}$ is measured. The experiment is repeated a number of times for different choices of local unitaries. From the statistics and the moments of the resulting probability distribution, one then aims to infer properties of the underlying quantum state. 

More formally, the quantities obtained in the experiment for a bipartite quantum state $\rho$ are those given in Eq.~(\ref{eq:mom}), where the integrals are evaluated with respect to the Haar measure over the unitary group $\mathcal U(d)$, the measured observable is denoted $\mathcal{M}$, and the moment of the corresponding probability distribution is denoted by~$t$.

In this paper, we are mainly concerned with two-qubit states, for which a complete generating set of 18 polynomial invariants has been characterized  before \cite{makhlin2002nonlocal}. Of these invariants, six are needed only to distinguish certain specific states by the signs of these invariants, thus we do not expect to extract relevant information in terms of entanglement or nonlocality from these. The remaining twelve invariants are of degree up to six. In order to express them properly, let us decompose the bipartite quantum state in terms of the Bloch representation, i.e., we write
\begin{align}
    \rho = \frac14\!\!\left[ \one\otimes \one + \vec{\alpha}\cdot \vec{\sigma} \otimes \one + \one \otimes \vec{\beta}\cdot\vec{\sigma} + \! \sum_{i,j=1}^3\! T_{ij} \sigma_i \otimes \sigma_j\right],
\end{align}
where $\sigma_{1,2,3}$ denote the usual Pauli matrices. Thus, the state is determined by its local Bloch vectors $\vec{\alpha}$ and $\vec{\beta}$, and the real correlation matrix $T$. In terms of these, the invariants can be expressed conveniently, and we give a complete list in Appendix~\ref{app:luinvs}. For our purposes, we will focus on the invariants  
$I_1 = \det(T)$, $I_2 = \trace(TT^T)$ and $I_3 = \trace(TT^\text{T}TT^\text{T})$. With the help of these three invariants, it is possible to decide whether the state can violate a CHSH-like Bell inequality. Furthermore, it is possible to bound the teleportation fidelity of the state.

Two of the invariants, including $I_1$, are special in the sense that they flip signs under partial transposition of $\rho$, whereas all other invariants do not. This has consequences on how to measure them: While the other invariants can be obtained from the statistics of product observables, $I_1$ and $I_{14}$ require non-product observables. In turn, they are linked to the entanglement of the state and can be used to obtain the entanglement measure of negativity of the state using randomized measurements  \cite{vidal2002computable}, see Appendix~\ref{app:luinvs} for more details.

\vspace{1em}
{\it Expressions for the LU invariants.---}
Let us now state explicitly how to measure the LU invariants in a randomized measurement scheme. To that end, recall that in order to observe the moments in Eq.~(\ref{eq:mom}), one has to choose an appropriate observable. Here, we show how to choose it in order to obtain the invariants $I_1$, $I_2$ and $I_3$.

As a first example, we explore the moments in case of the choice $\mathcal{M} = Z \otimes Z$. Note that choosing any other combination of Pauli matrices yields the same results, as they are related by local unitary rotations. For this choice, the first moment vanishes and we obtain as the second moment
\begin{align}
    \RR{2}{Z\otimes Z} &= \frac19 \trace(TT^T) = \frac19 I_2,
    \label{eq:R2ZZ}
\end{align}
where the occurring integrals can be solved using Weingarten calculus \cite{collins2022weingarten}.

Next, $t=3$ yields again zero (as well as any odd moment). For $t=4$, we obtain 
\begin{align}
    \RR{4}{Z\otimes Z} &= \frac{1}{75}[2\trace(TT^TTT^T) +\trace(TT^T)^2] \nonumber \\
    &= \frac1{75}[2I_3 + I_2^2],
    \label{eq:R4ZZ}
\end{align}
giving access to the invariant $I_3$.

Finally, as $I_1 = \det(T)$ flips sign under partial transposition, we consider the non-product observable $\mathcal{M}_{\det} = \sum_{i=1}^3 \sigma_i\otimes \sigma_i$ 
and $t=3$. Note that even though the observable is non-product, the moments can still 
be obtained by local measurements, as
the expectation value can be obtained from the three measurements
$X\otimes X$, $Y\otimes Y$, $ Z\otimes Z$ for a fixed choice of 
unitaries. The corresponding moment yields
\begin{align}
    \RR{3}{\mathcal{M}_{\det}}(\rho) = \det(T) = I_1.
    \label{eq:R3Adet}
\end{align}

This scheme is not limited to bipartite systems. Indeed, it is possible to measure a mixed-state variant of the Kempe invariant of three-qubit systems \cite{kempe1999multiparticle, barnum2001monotones}. We demonstrate this using the observable $\mathcal{M}_\text{Kempe} = Z\otimes Z\otimes \one + Z \otimes \one \otimes Z + \one \otimes Z\otimes Z$ in Appendix~\ref{app:luinvs}.
Before turning to the experimental implementation of the randomized measurement scheme, some statistical considerations are in order. For any fixed choice of local unitaries, multiple measurements are needed to obtain an estimate of the expectation value. Furthermore, a large number of  random unitaries have to be chosen. We denote the number of random unitary choices by $M$, and the number of measurements per choice to obtain the expectation value by $K$, such that the total number of measurements is given by $MK$.

Central for this scheme is the generation of Haar random unitaries. We certify 
the randomness of unitaries in our setup by calculating their so-called frame 
potential as detailed in Appendix~\ref{app:randomunitaries}. 

\begin{figure*}[!t]
    \centering
    \includegraphics[width=2.0\columnwidth]{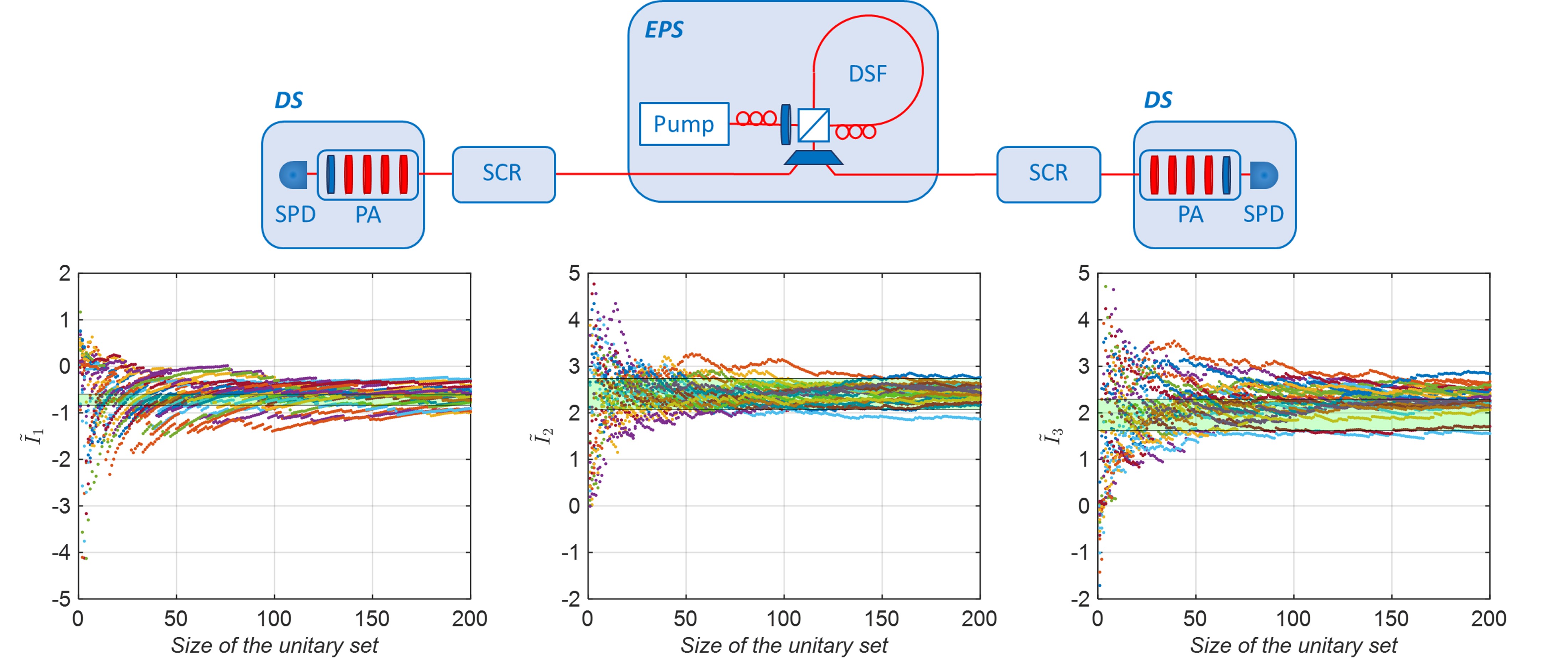}
    \caption{
    (a) Schematic diagram of the experimental setup for performing the randomized measurement protocol with polarization-entangled photon pairs. DS: detector station. DSF: dispersion-shifted fiber. EPS: entangled photon source. PA: polarization analyzer. SCR: polarization scrambler. SPD: single photon detector. 
    (b, c, d) Experimentally determined unbiased estimators for the invariants $I_{1}$, $I_{2}$, and $I_{3}$ for 25 different runs consisting of measurements with 200 different random unitaries. The black lines show the expected values of each parameter calculated from the density matrix of the experimental system.
    }
    \label{fig:exp_data}
\end{figure*}

\vspace{1em}
{\it Experimental methods.---}
We experimentally verify the functionality of the proposed randomized measurement 
method for the two-qubit case using polarization-entangled photon pairs.
A schematic of our experimental setup is shown in Fig. \ref{fig:exp_data}(a).
The entangled photon source (EPS) generates signal and idler photon pairs via four-wave mixing in a dispersion shifted fiber (DSF) \cite{fiorentino2002all}.
The DSF is pumped with a 50 MHz pulsed fiber laser centered at 1552.52 nm and is arranged in a Sagnac loop with a polarizing beam splitter (PBS) to entangle the signal and idler in polarization.
The photons are spectrally demultiplexed into 100 GHz-spaced channels on the International Telecommunication Union (ITU) grid after the Sagnac loop, resulting in photons with a temporal duration of about 15 ps \cite{wang2009robust, nucrypt}. 
For the experiment described here, ITU channels 27 (1555.75 nm) and 35 (1549.32 nm) are used.  
The source is tunable and typically outputs $\mu = 0.001 - 0.1$ pairs per pump pulse. 
Each photon is detected with gated InGaAs detectors with detection efficiencies of $\eta \sim 20\%$ and dark count probabilities of $ \sim 4 \times 10^{-5}$ per gate \cite{jones2017joint,jones2017situ}.

Given the polarization-entangled state generated by our source, we must implement random local unitary rotations in the form of random polarization state rotations. 
Therefore, we utilize the scrambling function of automated polarization controllers in order to apply random polarization rotations (for the remainder of the paper, a polarization controller operating in scrambling mode will be referred to as a polarization scrambler).

After verifying that the polarization scramblers can be used to apply sufficiently-random unitaries, we measured unbiased estimators (see Appendix~\ref{app:unbiased_estimators}) for the $I_{1}$, $I_{2}$, and $I_{3}$ invariants 
via Eqs.~(\ref{eq:R2ZZ})-(\ref{eq:R3Adet}).
Each polarization scrambler was set to rotate incident light to a random polarization state (therefore, acting as a random unitary), and coincidences were measured in different bases. To that end, we define for each of the two parties $i=1,2$ the local bases $\{\ket{H}_i, \ket{V}_i\}$ of horizontally and vertically polarized light, $\{\ket{D}_i, \ket{A}_i\}$ of diagonal and anti-diagonal polarized light and $\{\ket{L}_i, \ket{R}_i\}$ of left circular and right circular polarized light. Note that while we associate these bases to polarization states, the unitary invariance of the measured quantities allows us to choose any local bases, as long as they are rotated by $\frac\pi2$ on the Bloch sphere w.r.t.~each other. In particular, the bases for measuring photons 1 and 2 do not need to be aligned, i.e. $|H\rangle_{1}$ and $|H\rangle_{2}$ do not need to be equivalent on their respective Bloch spheres.
We then measured in each combination of these local bases repeatedly for $M=200$ different settings of the polarization scramblers, i.e. 200 different random unitaries were applied, where for each of these settings, $K\approx 1500$ repetitions were used to measure the expectation value.

The method to estimate $I_{1}$, $I_{2}$, and $I_{3}$ from finite measurement results is discussed in detail in Appendix~\ref{app:unbiased_estimators}. Although we collect measurement results in the $\{\ket{H}_i, \ket{V}_i\}$, $\{\ket{D}_i, \ket{A}_i\}$ and $\{\ket{L}_i, \ket{R}_i\}$ bases described above, the estimators for $I_{2}$ and $I_{3}$
only require projective measurements in a single joint basis. Therefore, those estimators are calculated using only a subset of the data, for example, the results for $|H\rangle_{1}|H\rangle_{2}$, $|H\rangle_{1}|V\rangle_{2}$, $|V\rangle_{1}|H\rangle_{2}$, and $|V\rangle_{1}|V\rangle_{2}$. On the other hand, the estimator for $I_{1}$ requires projective measurements in all three of the measured joint bases
After calculating the invariants, the experiment described above was repeated 25 different times to allow for a statistical analysis of the results.

The experimentally determined estimators of the $I_{1}$, $I_{2}$, and $I_{3}$ invariants for all 25 runs (each run is shown in a different color) are shown in Fig. \ref{fig:exp_data}(b), (c), and (d), respectively. 
The green band in all plots corresponds to the expected value of each invariant to allow for comparison with our method. The band represents the mean value plus or minus the standard deviation of each invariant calculated by performing quantum state tomography many times to characterize the state output by the EPS. A more-detailed description of how these expected values are calculated is found in Appendix~E3.
The experimentally determined invariants converge near the expected values, therefore validating our randomized measurement protocol.


\vspace{1em}
\vspace{1em}
{\it Applications to the detection of Bell nonlocality and teleportation fidelity.---}
The most straight-forward application is the evaluation of $I_2 = \trace(TT^\text{T})$, also known as the two-body sector length \cite{wyderka2020characterizing}. A quantum state is entangled if $I_{2}>1$, and the maximal value is $I_{2}=3$ for Bell states. Note that additional knowledge of $I_3=\trace(TT^\text{T}TT^\text{T})$ allows for the detection of many more entangled states \cite{imai2021bound}.

Combined knowledge of $I_1 = \det(T)$, $I_2 = \trace(TT^T)$ and $I_3 = \trace(TT^\text{T}TT^\text{T})$ is useful for completely determining if a state's nonlocality can be detected by a CHSH-like inequality:
Given the observable  \cite{verstraete2002entanglement}
\begin{align}
    \mathcal{B} = \sum_{i,j=1}^3 [a_i(c_j+d_j) + b_i(c_j-d_j)]\sigma_i \otimes \sigma_j,
\end{align}
where $\vec{a}$, $\vec{b}$, $\vec{c}$ and $\vec{d}$ are real, normalized vectors, its expectation value is bounded by 2 for local states. For a given quantum state, the maximum expectation value that one can observe by varying the vectors that define the observable is given by $2\sqrt{\lambda_1^2 + \lambda_2^2}$, where $\lambda_1$ and $\lambda_2$ are the two largest singular values of the correlation matrix $T$ \cite{horodecki1995violating}. Thus, the quantity 
\begin{align}\label{eq:chshviolation}
\text{CHSH}(\rho) = 2\sqrt{\lambda_1^2 + \lambda_2^2} - 2
\end{align}
measures the observable violation.

As the squares of the singular values of $T$ coincide with the eigenvalues of $TT^\text{T}$, we can obtain them by measuring the coefficients of the characteristic polynomial
\begin{multline}
    p_T(x) = x^3 - \trace(TT^\text{T})x^2- \\
-    \frac12[\trace(TT^\text{T}TT^\text{T})-\trace(TT^\text{T})^2]x-\det(T)^2,\label{eq:charpol}
\end{multline}
which are LU invariants, and calculating its roots. However, some care is needed to properly transfer statistical errors from finite statistics of the invariants to the roots of this polynomial; we explain the data analysis methods in Appendix~\ref{app:xiaodong}.

As a second figure of merit, we can decide whether a given two-qubit state is useful in a teleportation protocol. There, the maximal fidelity $f_\text{max}$ of the teleported state is given by \cite{horodecki1999general}
\begin{align}\label{eq:telfid}
    f_\text{max} = \frac{F_\text{max}d+1}{d+1},
\end{align}
where in our case $d=2$ and $F_\text{max}$ is the maximal overlap of the distributed state with the maximally entangled state $\ket{\phi^+} = \frac{1}{\sqrt{2}}(\ket{00}+\ket{11})$ under local operations and classical communication. As local unitary rotations constitute a subset of these, we can lower bound $F_\text{max}$ by optimizing over LUs instead, yielding \cite{guhne2021geometry}
\begin{align}\label{eq:maxfidelity}
    F_\text{max} \!\geq\! F_\text{max}^{U} \!:=\!  \frac14\max\{&1\!-\!\lambda_1\!-\!\lambda_2\!-\!\lambda_3, 1\!-\!\lambda_1\!+\!\lambda_2\!+\!\lambda_3,\nonumber\\
    &1\!+\!\lambda_1\!-\!\lambda_2\!+\!\lambda_3, 1\!+\!\lambda_1\!+\!\lambda_2\!-\!\lambda_3\}.
\end{align}
By examining the invariants $I_1$, $I_2$ and $I_3$, we can minimize 
$F_\text{max}^U$ over all singular values $\lambda_i$ which are compatible 
with the observed data, giving a lower bound on the teleportation 
fidelity of the prepared quantum state.

\vspace{1em}
{\it Results.---}
Using the methods described above and under the assumption that 25 repetitions of the experiment yields results which are well described by the Gaussian approximation, we extract the following experimental values for the invariants:
\begin{align}
    \det(T) &= -0.62\pm 0.15,\\
    \trace(TT^T) &= \phantom{-}2.41 \pm 0.15,\\
    \trace(TT^TTT^T) &= \phantom{-}2.21 \pm 0.21,
\end{align}
where the confidence regions correspond to $3\sigma$, i.e., 99.73\% confidence levels.
These values are all in agreement with the values determined from quantum state tomography (shown by the green bands in Fig. \ref{fig:exp_data}(b-d)): $I_{1} = -0.71\pm0.12$, $I_{2} = 2.41\pm0.34$, and $I_{3} = 1.95\pm0.34$ with $1\sigma$ confidence regions.

From these values, we obtain a potential CHSH violation of
\begin{align}
    \text{CHSH}(\rho) &\geq 0.46.
\end{align}
The confidence of this violation is given by $0.991 \approx 2.6\sigma$, as detailed in Appendix~\ref{app:xiaodong}. 
For comparison, the \textit{maximal} CHSH violation calculated from quantum state tomography is $\text{CHSH}_{\text{QST}} \leq 0.60\pm0.11$.
Note that the maximal observable value for a fully entangled state is given by $2\sqrt{2} - 2 \approx 0.83$.

Similarly, by requiring a higher confidence level of $5\sigma$ for the invariants, $\text{CHSH}(\rho) = 0.42$
with confidence $0.999998\approx 4.7\sigma$ can be obtained.

Using either method, our results clearly show that the randomized measurement protocol successfully determines that the state output by our EPS has the potential to violate a CHSH inequality.

For the teleportation fidelity, a confidence level of $3\sigma$ of the LU invariants yields a fidelity of at least
\begin{align}
    F_{\text{max}}^{U} &=0.88,
\end{align}
or, via Eq.~(\ref{eq:telfid}), $f_\text{max} = 0.92$, with a confidence level of $0.991 \approx 2.6\sigma$, 
By raising the confidence of the invariants to $5\sigma$, the lower bound decreases to $F_{\text{max}}^{U}=0.86$, or $f_\text{max} = 0.90$,
with confidence $0.999998 \approx 4.7\sigma$.
For comparison, the fidelity of the state calculated from tomography is $F_{\text{QST}} =0.90\pm0.08$, and the teleportation fidelity is $f_{\text{QST}} = 0.93\pm0.05$, confirming that the randomized measurement protocol accurately determines these parameters.

A detailed derivation of these values can be found in Appendix~\ref{app:dataanalysis}, where we also give values for these quantities without the assumption of Gaussian distribution, by using the Hoeffding inequality instead.

\vspace{1em}
{\it Conclusion.---}
We showed that any local unitary invariant characterizing the quantum 
correlations in quantum states of two or more particles can be directly 
measured using the moments of randomized measurements. We exemplified 
this for two-qubit states, where we showed how all relevant LU 
invariants can be inferred from randomized measurements of appropriately 
chosen observables. 
We proceeded to demonstrate the practicality of the introduced methods by conducting an experiment with entangled photon pairs leading to an efficient measurement of the LU invariants $I_1$, $I_2$ and $I_3$. The latter allowed us to directly certify important properties of the state, i.e., its Bell nonlocality as well as its usefulness for quantum teleportation. Furthermore, as a necessary by-product of our investigations, we devised methods allowing for a characterization of the degree of randomness of a set of experimentally implemented unitary transformations. 

We emphasize the simplicity of the presented scheme which allows to infer several important properties of the underlying quantum state through a number of randomly assorted measurements. For this reason, it will be an interesting direction of future research to extend the present explicit constructions for two-qubit states also to higher-dimensional systems which likely will find ample applications in quantum communication tasks. Also, it would be desirable to extend our approach to the characterization of nonlocal quantum channels and multiparticle quantum correlations such as multi-setting Bell nonlocality or spin squeezing.

\begin{acknowledgments}
{\it Acknowledgments.---}N.W.~acknowledges support by the QuantERA project QuICHE via the German Ministry of Education and Research (BMBF
Grant No.~16KIS1119K). A.K.~acknowledges funding from the  
Ministry of Economic
Affairs, Labour and Tourism Baden-Württemberg, under the
project QORA. S.I.~acknowledges support by the DAAD.
J.L.B.~acknowledges support from the House of Young Talents of the University of 
Siegen.
O.G. acknowledges support by the Deutsche Forschungsgemeinschaft (DFG, German Research Foundation, project numbers 447948357 and 440958198), the Sino-German Center for Research Promotion (Project M-0294), the ERC (Consolidator Grant 683107/TempoQ) and the German Ministry of Education and Research (Project QuKuK, BMBF Grant No.~16KIS1618K).
\end{acknowledgments}

\bibliographystyle{apsrev4-1}

\bibliography{cite}

\appendix

\onecolumngrid
\section{Two-qubit and three-qubit LU invariants}
\label{app:luinvs}

Let us start with the case of bipartite systems. We expand a two-qubit state $\rho$ in the Bloch basis as
\begin{align}
    \rho = \frac14\left[ \one\otimes \one + \vec{\alpha}\cdot \vec{\sigma} \otimes \one + \one \otimes \vec{\beta}\cdot\vec{\sigma} + \sum_{i,j=1}^3 T_{ij} \sigma_i \otimes \sigma_j\right],
\end{align}
such that it is determined by the local Bloch vectors $\vec{\alpha}$ and $\vec{\beta}$, and the correlation matrix $T$.

In terms of these, the Makhlin invariants read \cite{makhlin2002nonlocal}
\begin{align}
    & I_4 = \alpha^2, &&I_7 = \beta^2, && I_2 = \trace(TT^T), \nonumber\\
    \,\nonumber \\
    & I_{12} = \vec{\alpha}T\vec{\beta}, && I_1 = \det(T), &&\nonumber\\
    \,\nonumber \\
    & I_5 = [\vec{\alpha}T]^2, && I_8 = [T\vec{\beta}]^2, && I_3 = \trace(TT^TTT^T),\nonumber\\
    & I_{14} = \trace[(\star\vec{\alpha})T(\star \vec{\beta})^TT^T], && \nonumber\\
    \,\nonumber\\
    &I_{13} = \vec{\alpha}TT^TT\vec{\beta}, &&I_6 = [\vec{\alpha} T T^T]^2, &&I_9 = [T^TT\vec{\beta}]^2.
\end{align}
Here, the first row contains all invariants of degree two, 
the second row those of degree three, the third and fourth row 
the degree four invariants and the last row displays the degree five and six invariants. Furthermore, the Hodge star $\star$ maps a vector to a skew-symmetric matrix via $(\star \vec{\alpha})_{ij} = \sum_{k}\epsilon_{ijk}\alpha_k$. Also, we sometimes write scalar products of vectors as 
$I_{12} = \vec{\alpha}T\vec{\beta}$, 
instead of the more formal $I_{12} = \vec{\alpha}^T T\vec{\beta}$. 
The invariants $I_1$ and $I_{14}$ flip sign under partial transposition of $\rho$, while the others are invariant under this map.

Before we start, notice that we aim to measure the moments 
\begin{align}\label{eq:momappendix}
    \RR{t}{\mathcal{M}}(\rho) := \iint \text{d}U_A \text{d}U_B \{\trace[(U_A\otimes U_B) \rho (U_A^\dagger \otimes U_B^\dagger) \mathcal{M}]\}^t. 
\end{align}
As an observable, we choose $\mathcal{M} = (k_A\one + l_A Z) \otimes (k_B\one + l_B Z)$. Note that this is the most general choice for product observables, as all other choices can be obtained by local rotations, which are averaged out in the integral. 

\subsection{\texorpdfstring{The invariants $\alpha^2$, $\beta^2$ and $\trace(TT^T)$}{The invariants a**2, b**2 and Tr(TT**T)}}

It is clear that the result of Eq.~(\ref{eq:momappendix}) must be expressible by local invariants, and for $t=2$, one can readily check that
\begin{align}
    \RR{2}{Z\otimes\one}(\rho) &= \int \text{d}U_A \{\trace[\rho_A U_A Z U_A^\dagger ]\}^2 \nonumber\\
    &=\frac14 \sum_{i,j=1}^3 \alpha_i \alpha_j \underbrace{\int \text{d}U \trace[\sigma_i U Z U^\dagger]\trace[\sigma_j U Z U^\dagger]}_{\frac43\delta_{ij}} \nonumber\\
    &= \frac13 \alpha^2.
\end{align}
Here, the integral can be solved using Weingarten calculus \cite{collins2022weingarten}. 

Likewise, $\beta^2$ can be obtained by measuring $\RR{2}{\one\otimes Z}$. Next, setting $\mathcal{M} = Z\otimes  Z$, we obtain
\begin{align}
    \RR{2}{Z\otimes Z} &= \frac1{16} \sum_{ijkl=1}^3 T_{ij}T_{kl}\int\text{d}U_A \trace[\sigma_iU_A ZU_A^\dagger]\trace[\sigma_kU_AZU_A^\dagger]\int \text{d}U_B \trace[\sigma_jU_BZU_B^\dagger]\trace[\sigma_lU_BZU_B^\dagger]\nonumber \\
    &= \frac19 \trace(TT^T),
    \label{eq:sector_length_infinite}
\end{align}
thus, all degree-two invariants can be obtained in this way.

\subsection{\texorpdfstring{The invariants $\trace(TT^TTT^T)$, $\vec{\alpha}T\vec{\beta}$, $[\vec{\alpha}T]^2$ and $[T\vec{\beta}]^2$}{The invariants Tr(TT**TTT**T), aTb, [aT]**2 and [Tb]**2}}

Next, for $t=4$ and $\mathcal{M} = Z\otimes Z$, we obtain integrals like
\begin{align}
    \int \text{d}U \trace[\sigma_{i_1}UZU^\dagger]\trace[\sigma_{i_2}UZU^\dagger]\trace[\sigma_{i_3}UZU^\dagger]\trace[\sigma_{i_4}UZU^\dagger] = \frac{16}{15}[\delta_{i_1i_2}\delta_{i_3i_4} + \delta_{i_1i_3}\delta_{i_2i_4}+\delta_{i_1i_4}\delta_{i_2i_3}],    
\end{align}
leading to 
\begin{align}
    \RR{4}{Z\otimes Z} = \frac{1}{75}[2\trace(TT^TTT^T) +\trace(TT^T)^2],
\end{align}
giving access to invariant $I_3$. For the degree-three invariant $\vec{\alpha}T\vec{\beta}$; however, marginal terms are needed, i.e., $k_A\neq0\neq k_B$. Therefore, we set $k_A=k_B\equiv k$, $l_A = l_B \equiv l$, and $t=3$. We obtain

\begin{align}
    \RR{3}{(k\one + l  Z)^{\otimes 2}} = k^6 + k^4l^2[\alpha^2+\beta^2] + \frac13 k^2l^4[\trace(TT^T) + 2\vec{\alpha}T\vec{\beta}].
\end{align}
Thus, knowledge of the degree-two invariants together with choosing $k\neq0\neq l$, one can extract the degree-three invariant $I_{12} = \vec{\alpha}T\vec{\beta}$.

Finally, we use the same technique to obtain the degree-four invariants $[\vec{\alpha}T]^2$ and $[T\vec{\beta}]^2$:

\begin{align}
    \RR{4}{(k\one + l  Z)^{\otimes 2}} = k^8 &+ 2k^6l^2[\alpha^2+\beta^2] \nonumber\\
    &+ \frac{2}{3}k^4l^4[\frac3{10}(\alpha^4+\beta^4) + \alpha^2\beta^2 +\trace(TT^T) +4\vec{\alpha}T\vec{\beta}] \nonumber \\
    &+ \frac{2}{15}k^2l^6[(\alpha^2+\beta^2)\trace(TT^T) + 2( [\vec{\alpha}T]^2 + [T\vec{\beta}]^2)] \nonumber\\
    &+ \frac1{75} l^8[2\trace(TT^TTT^T) + \trace(TT^T)^2].
\end{align}
Thus, the combination $[\vec{\alpha}T]^2 + [T\vec{\beta}]^2$ can be obtained from this symmetric measurement. If, instead, one chooses $k_A\neq k_B$, $l_A \neq l_B$, the individual terms can also be measured.

\subsection{\texorpdfstring{The invariants $\vec{\alpha}TT^TT\vec{\beta}$, $[\vec{\alpha} T T^T]^2$, and $[T^TT\vec{\beta}]^2$}{The invariants aTT**TTb, [aTT**T]**2, and [T**TTb]**2}}

We find
\begin{align}
    \RR{5}{(k\one + l  Z)^{\otimes 2}}&=k^{10} + \frac{10}{3} k^8 l^2[\alpha^2+\beta^2] \nonumber\\
    &+ \frac{10}{3}k^6l^4[\frac{3}{10}(\alpha^4+\beta^4) + \alpha^2\beta^2 +\frac{1}{3} \trace(TT^T) +2\vec{\alpha}T\vec{\beta}] \nonumber\\
    &+ \frac{2}{3}k^4l^6[(\alpha^2+\beta^2)(\trace(TT^T)+2\vec{\alpha}T\vec{\beta}) + 2( [\vec{\alpha}T]^2 + [T\vec{\beta}]^2)] \nonumber\\
    &+ \frac1{15} k^2l^8[2\trace(TT^TTT^T)
    + \trace(TT^T)(\trace(TT^T)+4\vec{\alpha}T\vec{\beta})+8\vec{\alpha}TT^TT\vec{\beta}],
\end{align}
which allows us to extract $\vec{\alpha}TT^TT\vec{\beta}$.

Moreover, Weingarten calculus yields the useful expansion
\begin{multline}
    \int \text{d}U
    \trace[\sigma_{i_1}U  ZU^\dagger]
    \trace[\sigma_{i_2}U  ZU^\dagger]
    \trace[\sigma_{i_3}U  ZU^\dagger]
    \trace[\sigma_{i_4}U  ZU^\dagger]
    \trace[\sigma_{i_5}U  ZU^\dagger]
    \trace[\sigma_{i_6}U  ZU^\dagger]\\
    = \frac{64}{105}\{
     \delta_{i_1i_2} [\delta_{i_3i_4}\delta_{i_5i_6} +\delta_{i_3i_5}\delta_{i_4i_6}+\delta_{i_3i_6}\delta_{i_4i_5}]
    +\delta_{i_1i_3} [\delta_{i_2i_4}\delta_{i_5i_6} +\delta_{i_2i_5}\delta_{i_4i_6}+\delta_{i_2i_6}\delta_{i_4i_5}]\\
    +\delta_{i_1i_4} [\delta_{i_2i_3}\delta_{i_5i_6} +\delta_{i_2i_5}\delta_{i_3i_6}+\delta_{i_2i_6}\delta_{i_3i_5}]
    +\delta_{i_1i_5} [\delta_{i_2i_3}\delta_{i_4i_6} +\delta_{i_2i_4}\delta_{i_3i_6}+\delta_{i_2i_6}\delta_{i_3i_4}]\\
    +\delta_{i_1i_6} [\delta_{i_2i_3}\delta_{i_4i_5} +\delta_{i_2i_4}\delta_{i_3i_5}+\delta_{i_2i_5}\delta_{i_3i_4}]\},
\end{multline}
which can readily be used to calculate
\begin{align}
    \RR{6}{(k\one + l  Z)^{\otimes 2}}&=k^{12} + 5 k^{10} l^2[\alpha^2+\beta^2] \nonumber\\
    &+ \frac{1}{3}k^8l^4[9(\alpha^4+\beta^4) + 30\alpha^2\beta^2 +5\trace(TT^T) +40\vec{\alpha}T\vec{\beta}] \nonumber\\
    &+k^6l^6[(\alpha^2+\beta^2)(\alpha^2\beta^2+2\trace(TT^T)+8\vec{\alpha}T\vec{\beta})
    + 4( [\vec{\alpha}T]^2 + [T\vec{\beta}]^2)
    +\frac{1}{7}(\alpha^6+\beta^6)] \nonumber\\
    &+ \frac1{5} k^4l^8[2\trace(TT^TTT^T)
    + \trace(TT^T)(\trace(TT^T)+8\vec{\alpha}T\vec{\beta}
    +\frac{5}{7}(\alpha^4+\beta^4)+2\alpha^2 \beta^2)]\nonumber\\
    &+\frac1{5} k^4l^8[16\vec{\alpha}TT^TT\vec{\beta}
    +(\frac{20}{7}\alpha^2+4\beta^2)[\vec{\alpha}T]^2
    +(\frac{20}{7}\beta^2+4\alpha^2) [T\vec{\beta}]^2
    +8[\vec{\alpha}T\vec{\beta}]^2]\nonumber\\
    &+\frac{1}{35}k^2l^{10}[(\alpha^2+\beta^2)(2\trace(TT^TTT^T)
    + \trace(TT^T)^2)+4([\vec{\alpha}T]^2 + [T\vec{\beta}]^2)\trace(TT^T)]\nonumber\\
    &+\frac{8}{35}k^2l^{10}[[\vec{\alpha} T T^T]^2 + [T^TT\vec{\beta}]^2]\nonumber\\
    &+\frac{1}{735}l^{12}[
    8\trace(TT^TTT^TTT^T)+\trace(TT^T)(6\trace(TT^TTT^T)+\trace(TT^T)^3)].
\end{align}
This expression yields $[\vec{\alpha} T T^T]^2$, and $[T^TT\vec{\beta}]^2$.

\subsection{Invariants from non-product observables}\label{app:nonprod}

There are two invariants left, namely $I_1 = \det(T)$ and $I_{14} = \trace[(\star\vec{\alpha})T(\star \vec{\beta})^TT^T]$. 
Before we continue, note that for the moments of the partially transposed state $\rho^{T_B} = (\text{id} \otimes \hat{T})(\rho)$, where $\hat{T}$ denotes the usual transposition map, it holds that
\begin{align}
\RR{t}{A\otimes B}(\rho^{T_B}) &:= \iint \text{d}U_A \text{d}U_B \{\trace[(U_A\otimes U_B) \rho^{T_B} (U_A^\dagger \otimes U_B^\dagger) (A\otimes B)]\}^t \nonumber\\
 &= \iint \text{d}U_A \text{d}U_B \{\trace[(U_A\otimes U_B) \rho (U_A^\dagger \otimes U_B^\dagger) (A\otimes B^T)]\}^t\nonumber\\
 &=\RR{t}{A\otimes B^T}(\rho).
\end{align}
However, if we write for qubits $B = k\one + l_X  X + l_Y  Y + l_Z  Z$, then $B^T = k\one + l_X  X - l_Y  Y + l_Z  Z$, which is related to $B$ via a simple unitary rotation. Therefore, $\RR{t}{A\otimes B}(\rho^{T_B}) = \RR{t}{A\otimes B}(\rho)$, as long as the observable is product. As noted before, however, invariants $I_1$ and $I_{14}$ flip sign under partial transposition, and can therefore not occur in these product moments.

To circumvent this problem, we consider instead non-product observables. We start by using $\mathcal{M}_{\det} = \sum_{i=1}^3 \sigma_i\otimes \sigma_i$ and $t=3$. Note that even though the observable is non-product, the moments can still be obtained by local measurements, as
\begin{align}
    \RR{t}{\mathcal{M}_{\det}} = \iint \text{d}U_A \text{d}U_B \{\trace[(U_A\otimes U_B) \rho (U_A^\dagger \otimes U_B^\dagger) \sum_{i}\sigma_i \otimes \sigma_i]\}^t,
\end{align}
and the expectation value can be obtained from the three measurements $  X\otimes   X$, $  Y\otimes  Y$, $  Z\otimes  Z$ for a fixed choice of unitaries.

We now evaluate explicitly
\begin{multline}
    \RR{3}{\mathcal{M}_{\det}}(\rho) = \iint \text{d}U_A \text{d}U_B\left[\sum_i \trace(\rho_U \sigma_i\otimes\sigma_i)^3 + 3\sum_{i\neq j}\trace(\rho_U \sigma_i\otimes\sigma_i)^2 \trace(\rho_U \sigma_j\otimes \sigma_j) \right.+\\
    \left.\phantom{\sum_i} + 6\trace(\rho_U   X\otimes   X)\trace(\rho_U   Y\otimes   Y)\trace(\rho_U   Z\otimes   Z) \right].
\end{multline}
Here, we used the abbreviation $\rho_U := (U_A\otimes U_B)\rho (U_A^\dagger \otimes U_B^\dagger)$.

Using Weingarten calculus again, one can quickly see that only the last term is non-vanishing with

\begin{align}
\int \text{d}U\trace(\sigma_{i}U  X U^\dagger)\trace(\sigma_{j}U  Y U^\dagger)\trace(\sigma_{k}U  Z U^\dagger) = \frac43\epsilon_{ijk}.
\end{align}
Thus,
\begin{align}
    \RR{3}{\mathcal{M}_{\det}}(\rho) &= 6\iint \text{d}U_A \text{d}U_B\trace(\rho_U   X\otimes   X)\trace(\rho_U   Y\otimes   Y)\trace(\rho_U   Z\otimes   Z) \nonumber\\
    &=\frac{6\cdot 4\cdot 4}{4^3 \cdot3\cdot 3}\sum_{i_1i_2i_3j_1j_2j_3} T_{i_1j_1}T_{i_2j_2}T_{i_3j_3}\epsilon_{i_1i_2i_3}\epsilon_{j_1j_2j_3} \nonumber\\
    &= \det(T) = I_1.
\end{align}

For invariant $I_{14}$, we need an expression of order four. We choose $\mathcal{M}_\text{Hodge} = \one\otimes   X +   X \otimes \one +   Y\otimes   Z +   Z\otimes   Y$. To see the moment distribution of this operator, we can apply the unitary average on it. To that end, we write
\begin{align}
    \RR{4}{\mathcal{M}} &= \iint \text{d}U_A \text{d}U_B \trace(\rho_U^{\otimes 4} \mathcal{M}^{\otimes 4}) \nonumber\\
    &= \iint \text{d}U_A \text{d}U_B \trace[\rho^{\otimes 4} (U_A\otimes U_B \mathcal{M} U_A^\dagger \otimes U_B^\dagger)^{\otimes 4}] \nonumber\\
    &= \trace\left[\rho^{\otimes 4} \iint \text{d}U_A \text{d}U_B (U_A\otimes U_B \mathcal{M} U_A^\dagger \otimes U_B^\dagger)^{\otimes 4}\right]. 
\end{align}
Thus, we can  twirl the operator instead of the state. To evaluate this, we first rearrange the parties in the four-copy space as $A_1, A_2, A_3, A_4, B_1, B_2, B_3, B_4$. Then, before the averaging, we get for $\mathcal{M}_\text{Hodge}$
\begin{align}\label{eq:hodgetwirl}
    \iint \text{d}U_A \text{d}U_B U_A^{\otimes 4} U_B^{\otimes 4} & (\one\one\one\one \otimes XXXX + XXXX\otimes \one\one\one\one + \one\one XX\otimes XX\one\one + \text{perms.} + \nonumber \\
    &+\one\one YY\otimes XXZZ + \text{perms.} + \one\one ZZ\otimes XXYY + \text{perms.} +\nonumber  \\
    &+XXZZ\otimes \one\one YY + \text{perms.} + XXYY \otimes \one\one ZZ+ \text{perms.} +\nonumber  \\
    &+YYZZ\otimes ZZYY + \text{perms.} + YYYY\otimes ZZZZ + ZZZZ \otimes YYYY + \nonumber \\
    &+\one XYZ \otimes X\one ZY + \text{perms.} + \ldots)(U_A^\dagger)^{\otimes 4}(U_B^\dagger)^{\otimes 4}.
\end{align}
Here, perms. denotes all permutations among the A and B parties of the preceding term. For instance, $\one \one XX\otimes XX\one\one + \text{perms.} = \one \one XX\otimes XX\one\one + \one X \one X\otimes X\one X\one+\one XX\one \otimes X \one \one X + X\one \one X \otimes \one XX\one + X\one X \one \otimes \one X \one X + XX \one \one \otimes \one \one XX$, yielding six terms in total. Note that we have already omitted all those terms that yield zero after applying the integral. 

In the following, we sloppily concentrate on those yielding invariant $I_{14}$: These are the terms in the last row of Eq.~(\ref{eq:hodgetwirl}), where we use
\begin{align}
    \int \text{d}U U^{\otimes 4} \one X Y Z (U^\dagger)^{ \otimes 4}=\sum_{\pi\in S_3} \operatorname{sgn(\pi)} \one \sigma_{\pi_1}\sigma_{\pi_2}\sigma_{\pi_3}.
\end{align}
This yields terms in the fourth moment of the following form:
\begin{align}
    \RR{4}{\mathcal{M}_{\text{Hodge}}}(\rho) &= \ldots + 24\iint \text{d}U_A \text{d}U_B \trace(\rho_U \one\otimes  X)\trace(\rho_U   X\otimes\one)\trace(\rho_U   Y\otimes  Z)\trace(\rho_U   Z\otimes  Y) \nonumber\\
    &= \ldots + \frac{24}{4^4}\sum_{i_2i_3i_4j_1j_3j_4}\beta_{j_1}\alpha_{i_2}T_{i_3j_3}T_{i_4j_4}\times \nonumber\\
    &\phantom{=\ldots + 2} \times \int \text{d}U_A \trace(\sigma_{i_2}U_A  XU_A^\dagger)\trace(\sigma_{i_3}U_A  YU_A^\dagger) \trace(\sigma_{i_4}U_A  ZU_A^\dagger) \times \nonumber\\
    &\phantom{=\ldots + 2} \times \int\text{d}U_B \trace(\sigma_{j_1}U_B  XU_B^\dagger)\trace(\sigma_{j_3}U_B  ZU_B^\dagger)\trace(\sigma_{j_4}U_B  YU_B^\dagger) \nonumber\\
    &=\ldots - \frac{24\cdot4\cdot4}{4^4\cdot 3\cdot 3}\sum_{i_2i_3i_4j_1j_3j_4}\beta_{j_1}\alpha_{i_2}T_{i_3j_3}T_{i_4j_4} \epsilon_{i_2i_3i_4}\epsilon_{j_1j_3j_4}\nonumber\\
    &=\ldots-\frac16\sum_{ijklmn}\epsilon_{ijk}\epsilon_{lmn} \alpha_i \beta_l T_{jm}T_{kn} = \ldots - \frac16\trace[(\star\vec{\alpha})T(\star \vec{\beta})^TT^T].
\end{align}
Finally, in order to get rid of the undetermined contributions, note that measuring $\mathcal{M}_\text{Hodge}^\prime = \one\otimes   X +   X \otimes \one +   Y\otimes   Z -   Z\otimes   Y$ instead, the only term changing sign in Eq.~(\ref{eq:hodgetwirl}) is the Hodge term. Thus, the difference of the measured fourth moment of $\mathcal{M}_\text{Hodge}$ and $\mathcal{M}_\text{Hodge}^\prime$ is equal to one third of the invariant in question.

\subsection{Expressing the invariants via fidelities}

Instead of measuring expectation values of abstract operators, we can instead think of these as average fidelities. Indeed, coming back to invariant $I_1 = \det(T)$, we can  make use of
\begin{align}
    \mathcal{M}_\text{det} = 4\ket{\psi^-}\bra{\psi^-} - \one\otimes\one, 
\end{align}
thus, we can extract the determinant from \begin{align}
\RR{4}{\ket{\psi^-}\!\bra{\psi^-}}(\rho) = \iint \text{d}U_A \text{d}U_B \braket{\psi^-|\rho_U|\psi^-}^4,
\end{align}
where $\rho_U = U_A\otimes U_B \rho U_A^\dagger \otimes U_B^\dagger$.

For $\mathcal{M}_\text{Hodge}$, we can use similar tricks. By noticing that
\begin{align}
    \ket{\nu_X} &= \frac12(1, e^{i\pi/4}, e^{i\pi/4}, 1)^T
\end{align}
can be expanded as
\begin{align}
    \ket{\nu_X} \bra{\nu_X}&=\frac14[\one \otimes \one + \frac1{\sqrt{2}}(\one \otimes X + X \otimes \one + Y\otimes Z + Z \otimes Y) + X\otimes X],
\end{align}
we see that 
\begin{align}
    \frac1{\sqrt{2}}\mathcal{M}_\text{Hodge} = 4\ket{\nu_X}\bra{\nu_X} - \one\otimes\one - X\otimes X,
\end{align}
which allows to extract $I_{14}$ in terms of the fourth moment of the overlap between $\rho$ and $\ket{\nu_X}$.
Note that, due to local unitary invariance, we can instead also use 
\begin{align}
    \ket{\nu_Z} &= (a,0,0,\sqrt{1-a^2})^T
\end{align}
with $a=\sqrt{\frac{1+\sqrt{2}}{2\sqrt{2}}} = \cos(\pi/8)$. To be more precise, measuring 
\begin{align}
\RR{4}{\ket{\nu_Z}\!\bra{\nu_Z}}(\rho) &= \iint \text{d}U_A \text{d}U_B \braket{\nu_Z|\rho_U|\nu_Z}^4 \nonumber\\
&=\frac{1}{75\cdot 2^{10}}\left[ 300(1-\alpha^2 - \beta^2) +400\trace(TT^T) - 600\det(T)+400 \vec{\alpha}T\vec{\beta}  \right. \nonumber \\
& \phantom{-----}+15(\alpha^4 + \beta^4)+50\alpha^2\beta^2+60(\alpha^2+\beta^2)\trace(TT^T) \nonumber\\
& \phantom{-----}\left.+20[(\vec{\alpha}T)^2 +(T\vec{\beta})^2] -23\trace(TT^TTT^T)+51\trace(TT^T)^2-50I_{14}\right].
\end{align}

\subsection{Quantification of negativity}\label{app:negativity}
A typical entanglement measure in two-qubit systems is the negativity \cite{vidal2002computable}, which is defined as
\begin{align}
    N(\rho) = -2\min\{0,\mu(\rho^{T_B})\},
\end{align}
where
$\mu(\rho^{T_B})$ is the minimal eigenvalue of the $\rho^{T_B}$, and we note that in the case of two entangled  qubits,
exactly one eigenvalue is negative 
\cite{sanpera1998local}.

With the help of Newton's identities, this eigenvalue can be calculated by the moments given by $p_k=\trace[(\rho^{T_B})^k]$.
In fact, it has been shown that the negativity can be obtained by solving the following fourth degree polynomial for $N$ \cite{bartkiewicz2015quantifying}:
\begin{align}
    48\det(\rho^{T_B})+3N^4+6N^3-6N^2(p_2-1)-4N(3p_2-2p_3-1)=0.
\end{align}
We remark that the determinant $\det(\rho^{T_B})$ can be rewritten in terms of the moments $p_k$ via \cite{augusiak2008universal}
\begin{align}
    \det(\rho^{T_B})
    = \frac{1}{24}
    (1-6p_4+8p_3+3p_2^2-6p_2).
\end{align}
Therefore, to quantify the negativity, it is sufficient to use known relations between the $p_k$ and LU invariants  \cite{bartkiewicz2015method}:
\begin{align}
    p_2 &= \frac{1}{4}(1+x_1),\\
    p_3 &= \frac{1}{16}(1+3x_1+6x_2),\\
    p_4 &= \frac{1}{64}(1+6x_1+24x_2+x_1^2+2x_3+4x_4),
\end{align}
where
\begin{align}
    x_1& = I_2+I_4+I_7,\\
    x_2& = I_1+I_{12} ,\\
    x_3& = I_2^2 -I_3,\\
    x_4& = I_5+I_8+I_{14}+I_4 I_7.
\end{align}

\subsection{The Kempe invariant of three-qubit systems}\label{app:kempe}

Let us conclude with an example of a three-qubit invariant. Analogously to the Bloch decomposition of two-qubit states, we can expand a three-qubit systems as

\begin{align}
    \rho &= \frac18\left[\vphantom{\sum_{i=1}^3}\one \otimes \one \otimes \one + \vec{\alpha}\cdot\vec{\sigma} \otimes \one \otimes \one +\one \otimes \vec{\beta}\cdot\vec{\sigma} \otimes \one +\one \otimes \one \otimes \vec{\gamma}\cdot\vec{\sigma} \right.+\nonumber\\
    &\phantom{\frac18[}+\sum_{i,j=1}^3 T^{AB}_{ij} \sigma_i \otimes \sigma_j \otimes \one + T^{AC}_{ij} \sigma_i \otimes  \one \otimes \sigma_j+ T^{BC}_{ij} \one \otimes \sigma_i \otimes \sigma_j +\nonumber\\
    &\phantom{\frac18[}+\left.\sum_{i,j,k=1}^3 W_{ijk} \sigma_i\otimes \sigma_j\otimes \sigma_k\right].
\end{align}

The Kempe invariant was originally defined for pure three-qubit states to distinguish states that have coinciding invariants of their two-qubit marginals \cite{kempe1999multiparticle}. It can be extended to mixed states, one such extension being \cite{barnum2001monotones}
\begin{align}
    I_\text{Kempe} = \trace[(\rho_{AB}\otimes \one_C)(\rho_{AC}\otimes \one_B)(\rho_{BC}\otimes \one_A)],
\end{align}
where $\rho_{AB} = \trace_C(\rho)$ denotes the marginal state of parties $A$ and $B$ (and like wise for $\rho_{AC}$ and $\rho_{BC}$). In terms of the Bloch representation, it can be expressed as
\begin{align}
    I_{\text{Kempe}} = \frac18\left[1 + \alpha^2 + \beta^2 + \gamma^2 + \vec{\alpha}T^{AB}\vec{\beta} + \vec{\alpha}T^{AC}\vec{\gamma} + \vec{\beta}T^{BC}\vec{\gamma} + \trace(T^{AB}T^{BC}T^{CA})\right].
\end{align}
Here, we defined $T^{CA} = (T^{AC})^T$. Note that all but the last term are actually invariants of the bipartite marginals of the state, which we can measure using the methods developed before. Only the term $\trace(T^{AB}T^{BC}T^{CA})$ requires a proper three-qubit observable to be measured. To that end, we consider the observable $\mathcal{M}_\text{Kempe} = Z\otimes Z \otimes \one + Z \otimes \one \otimes Z + \one \otimes Z \otimes Z$. The third moment reads
\begin{align}
    \RR{3}{\mathcal{M}_\text{Kempe}}(\rho) &= \iiint \text{d}U_A \text{d}U_B \text{d}U_C \{\trace[(U_A\otimes U_B\otimes U_C) \rho (U_A^\dagger \otimes U_B^\dagger \otimes U_C^\dagger) \mathcal{M}_\text{Kempe}\}^3
\end{align}
In order to evaluate it using Weingarten calculus, we observe that for all $i,j,k\in\{0,\ldots,3\}$
\begin{align}
\int \text{d}U\trace(\sigma_{i}U  Z U^\dagger)\trace(\sigma_{j}U \one U^\dagger)\trace(\sigma_{k}U \one U^\dagger) &= 0,\\
\int \text{d}U\trace(\sigma_{i}U Z U^\dagger)\trace(\sigma_{j}U Z U^\dagger)\trace(\sigma_{k}U \one U^\dagger) &= \frac83\delta_{ij}(1-\delta_{i0})\delta_{k0},
\end{align}
such that 
\begin{align}
    \RR{3}{\mathcal{M}_\text{Kempe}}(\rho) &= \frac6{8^3}\frac{8^3}{3^3}\sum_{i_1,i_2,i_3=0}^3\sum_{j_1,j_2,j_3}^3\sum_{k_1,k_2,k_3=0}^3\trace(\rho \sigma_{i_1}\otimes \sigma_{i_2}\otimes \sigma_{i_3})\trace(\rho \sigma_{j_1}\otimes \sigma_{j_2}\otimes \sigma_{j_3})\trace(\rho \sigma_{k_1}\otimes \sigma_{k_2}\otimes \sigma_{k_3}) \times \nonumber\\
    &\hspace{17em}\times[\delta_{i_1j_1}(1-\delta_{i_10})\delta_{k_10}][\delta_{i_2k_2}(1-\delta_{i_20})\delta_{j_20}][\delta_{j_3k_3}(1-\delta_{j_30})\delta_{i_30}] \nonumber \\
    &= \frac29 \trace(T^{AB}T^{BC}T^{CA}),
\end{align}
giving access to the Kempe invariant.

\section{Randomness of unitary gates}
\label{app:randomunitaries}

\subsection{Frame potential}
A major concern in the experimental setup is the generation of random local unitary rotations. Since the framework of randomized measurements asserts that the unitaries must be distributed according to the Haar measure, it is important to check that the experimentally applied unitaries are indeed Haar random.

In order to check this, we resort to the so-called frame potential and unitary designs \cite{gross2007evenly, scott2008optimizing,  hunter2019unitary}. A set $\mathcal{U}=\{U_i\}_{i=1}^N$ of $N$  unitary gates is called a unitary $t$-design, if 
\begin{equation}
    \intop p(U)\,\text{d}U = \frac1N\sum_{i=1}^N p(U_i)
\end{equation}
for each polynomial $p$ of degree $t$ in the entries of the unitaries. That is, one can replace the integration over the Haar measure by averaging over a finite set of (carefully chosen) unitaries. The number of elements of such designs, which are known to exist for each dimension $d$ and order $t$, naturally grows with the order $t$ of the polynomial one wishes to average over. As the criterion we want to check involves invariants of order up to four, we are interested mainly in four-designs.

In order to check whether a given set $\mathcal{U}$ of $N$ unitaries constitutes a four-design, one can use the frame potential. It is defined for a set of unitaries via
\begin{equation}\label{eq:framepotential}
    F_t(\mathcal{U}) := \frac1{N^2} \sum_{U,V\in \mathcal{U}} \vert\trace(UV^\dagger)\vert^{2t}.
\end{equation}
Interestingly, it is minimized by unitary $t$-designs as well as in the limit of infinite sets of Haar random unitaries. 
The minimal value for $d=2$, i.e., qubit systems, is given 
by  $F_t^\text{Haar} = \frac{(2t)!}{t!(t+1)!}$ \cite{gessel1990symmetric}.

Now, if in an experiment, $N\rightarrow \infty$ unitaries are drawn, one can evaluate the frame potential for $t=4$ and compare it to the minimum, and if they match, one can be sure to have drawn them correctly (at least for the purposes of the task at hand; otherwise, $t$ has to be adjusted accordingly).
However, if $N$ is a finite number, one would expect a deviation from the minimal number, even if the distribution is perfectly Haar random. To quantify this, we introduce the excess quantity
\begin{align}
    G_t(U_1,\ldots,U_N) := \frac{F_t(U_1,\ldots,U_N) }{F_t^\text{Haar}}.
\end{align} Then, the expected excess $\mathbb{E}_{U_1,\ldots,U_N}\big[G_t(U_1,\ldots,U_N)\big]$ is given by
\begin{align}
    \mathbb{E}_{U_1,\ldots,U_N}\big[G_t(U_1,\ldots,U_N)\big]  &= \intop\text{d}U_1\ldots\text{d}U_N \frac{F_t(U_1,\ldots,U_n) }{F_t^\text{Haar}}\nonumber\\
    &= \frac1{N^2F_t^\text{Haar}}\sum_{i,j=1}^N \intop \text{d}U_1\ldots \text{d}U_N \vert\trace(U_i U_j^\dagger)\vert^{2t}\nonumber\\
    &= \frac1{N^2F_t^\text{Haar}}\left[Nd^{2t} + \sum_{i\neq j}\intop  \text{d}U\text{d}V  \vert\trace(UV^\dagger)\vert^{2t}\right] \nonumber\\
    &= \frac1{N^2F_t^\text{Haar}}\left[Nd^{2t} + N(N-1)F_t^\text{Haar}\right]\nonumber\\
    &= \frac{d^{2t}}{NF_t^\text{Haar}} + \frac{N-1}{N},\label{eq:expft}
\end{align}
where the integral in the third line can either be solved via Weingarten calculus, or by noting that it coincides precisely with the value of the frame potential for Haar random unitaries, i.e., it yields its minimum $F_t^\text{Haar}$. 
From Eq.~(\ref{eq:expft}) it can be directly seen that the expectation value approaches one only in the limit $N\rightarrow \infty$, and otherwise is larger than one. In order to quantify the probability to observe an excess of this expectation value larger than $\delta$, we make use of the Cantelli inequality, stating that for a set of randomly drawn unitaries $\{U_1, \ldots, U_N\}$ \cite{ghosh2002probability}, 
\begin{align}
    p(G_t(\{U_i\}) \geq \mathbb{E}(G_t) + \delta) \leq \frac{\operatorname{Var}(G_t)}{\delta^2 + \operatorname{Var}(G_t)}
\end{align}
with $\operatorname{Var}(G_t) = \mathbb{E}(G_t^2) - \mathbb{E}(G_t)^2$, and the abbreviation $\mathbb{E}(F) = \mathbb{E}_{U_1,\ldots,U_N}F(U_1,\ldots,U_N)$.
We calculate
\begin{align}
    \mathbb{E}(G_t^2) = \frac1{N^4 (F_t^\text{Haar})^2} \sum_{ijkl=1}^N \intop \text{d}U_1\ldots\text{d}U_N \vert\trace(U_iU_j^\dagger)\vert^{2t}\vert\trace(U_kU_l^\dagger)\vert^{2t}.
\end{align}
In order to proceed, several cases concerning the summation variables have to be distinguished:
\begin{align}
    (A)~i=j=k=l, && (B)~i=j=k\neq l, && (C)~i=j\neq k=l, \nonumber\\
    (D)~i=k\neq j=l, && (E)~i=j\neq k\neq l, && (F)~i=k\neq j\neq l,\nonumber\\
    (G)~i\neq j \neq k \neq l.
\end{align}
(Note, that here, $j\neq k \neq l$ is to be understood to also imply $j\neq l$). Each of the cases can be solved individually, but occurs multiple times. We go through the expressions one at a time.
\begin{align}
    (A) &= \frac{d^{4t}}{N^4 (F_t^\text{Haar})^2} = (C), \\
    (B) &= \frac1{N^4 (F_t^\text{Haar})^2} d^{2t} \intop \text{d}U \text{d}V \vert \trace(UV^\dagger)\vert^{2t} = \frac{d^{2t}}{N^4 F_t^\text{Haar}} = (E), \\
    (D) &= \frac1{N^4 (F_t^\text{Haar})^2} \intop \text{d}U \text{d}V \vert \trace(UV^\dagger)\vert^{4t} = \frac{F_{2t}^\text{Haar}}{N^4 (F_t^\text{Haar})^2},\\
    (F) &= \frac1{N^4 (F_t^\text{Haar})^2} \intop \text{d}U \text{d}V \text{d}W \vert \trace(UV^\dagger)\vert^{2t}\vert \trace(UW^\dagger)\vert^{2t} \nonumber\\
    &= \frac1{N^4 (F_t^\text{Haar})^2} \intop \text{d}U \left(\intop \text{d}V \vert \trace(UV^\dagger)\vert ^{2t}\right)^2 \nonumber\\
    &= \frac1{N^4 (F_t^\text{Haar})^2} \intop \text{d}U \left(\intop \text{d}(VU^\dagger) \vert \trace(UV^\dagger)\vert ^{2t}\right)^2 \nonumber\\
    &= \frac1{N^4 (F_t^\text{Haar})^2} \intop \text{d}U (F_t^\text{Haar})^2 =\frac1{N^4}, \\
    (G) &= \frac1{N^4 (F_t^\text{Haar})^2} \intop \text{d}U \text{d}V \text{d}W \text{d}X \vert \trace(UV^\dagger)\vert^{2t}\vert \trace(WX^\dagger)\vert^{2t} = \frac1{N^4} = (F).
\end{align}
Note that each term occurs multiple times, depending on the number of configurations. Thus, we collect all of them to obtain
\begin{align}
    \mathbb{E}(G_t^2) &= N(A) + 4N(N-1)(B) + N(N-1)(C) + 2N(N-1)(D) +\nonumber\\ 
    &\phantom{=}+ 2N(N-1)(N-2)(E) + 4N(N-1)(N-2)(F) +\nonumber \\
    &\phantom{=}+N(N-1)(N-2)(N-3)(G)
\end{align}
Inserting the results and subtracting $\mathbb{E}(G_t)^2$ yields
\begin{align}
    \operatorname{Var}(G_t) = \frac{2N(N-1)}{N^4}\left[\frac{F_{2t}^\text{Haar}}{(F_t^\text{Haar})^2}-1\right]
\end{align}

\begin{figure}
    \centering
    \includegraphics[width=0.7\columnwidth]{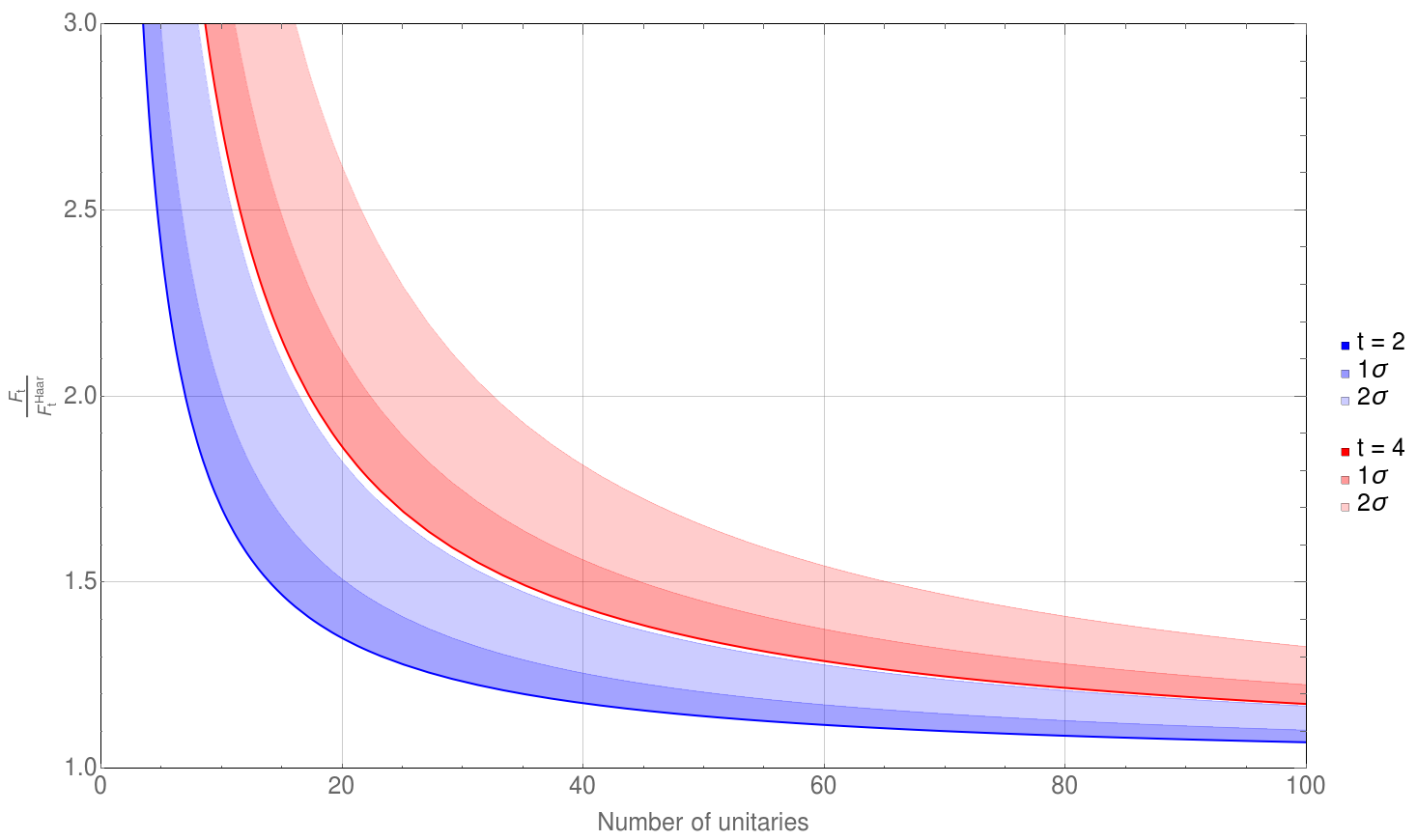}
    \caption{The expected value of the frame potential in Eq.~(\ref{eq:framepotential}) for Haar-randomly drawn sets of unitaries as a function of the set size for $t=2$ (blue) and $t=4$ (red). The colored areas above the curve denote the $1\sigma$ and $2\sigma$ confidence regions that are used to check whether the generated unitaries are compatible with the assumption of Haar-randomness. Note that values below the curves are attainable, but our main concern are sets of points whose frame potential values are higher than expected. Thus, we give one-sided confidence regions here.}
    \label{fig:framepotential}
\end{figure}

Plugging this into Cantelli's inequality, we obtain the plot in Fig.~\ref{fig:framepotential}. It is to be understood as follows: Setting the right hand side to 10\%, we obtain an error band that allows us to conclude that for a set of correctly drawn unitaries of certain size, the value of the frame potential will lie with 90\% probability within the marked area. Conversely, if one observes a value above that threshold, then with high probability the process was not Haar random.

\subsection{Spherical frames}
In the experiment at hand, we characterize the randomly drawn unitaries by applying them to the fixed input state $\ket{0}$ and measuring the resulting state. Instead of yielding the random unitaries, this yields instead a set of random states $\{\ket{\psi_i}\}_{i=1}^N$, and the underlying unitary can only be reconstructed up to a relative phase. 

We can remedy this by analyzing the randomness of the output states instead. A unitary $t$-design generates a spherical $t$-design by applying the unitaries to some fixed state. Analogously to the case of unitary designs, a spherical $t$-design is a set of vectors $\mathcal{S} = \{\ket{\psi_i}\}_{i=1}^N$, such that
\begin{align}
    \intop p(\ket{\psi}) \text{d}\!\ket{\psi} = \frac1N\sum_{i=1}^N p(\ket{\psi_i})
\end{align}
for all polynomials of degree $t$ or less in the entries of the vector $\ket{\psi}$. Also in this case, one can define the (spherical) frame potential,
\begin{align}
    \tilde{F}_t(\mathcal{S}) := \frac1{N^2} \sum_{\ket{\phi}, \ket{\psi} \in \mathcal{S}} \vert\braket{\psi|\phi}\vert^{2t},
\end{align}
which is minimized iff $\mathcal{S}$ is a spherical $t$-design. In contrast to the unitary case, the minimal value is given by  $\tilde{F}_t^\text{Haar} = \frac{t!(d-1)!}{(t+d-1)!}$ \cite{renes2004frames}. 

In complete analogy to the unitary case, we can define, for a given set of vectors $\mathcal{S}$, the excess quantity $\tilde{G}_t(\mathcal{S}) = \frac{\tilde{F}_t(\mathcal{S})}{\tilde{F}_t^\text{Haar}}$.

Calculating expectation values and confidence bounds for finite $N$ follows exactly the same steps as in the unitary case and yields
\begin{align}\label{eq:expGt}
    \mathbb{E}(\tilde{G}_t) &= \frac{1}{N\tilde{F}_t^\text{Haar}} + \frac{N-1}N,\\
    \operatorname{Var}(\tilde{G}_t) &= \frac{2N(N-1)}{N^4}\left[\frac{\tilde{F}_{2t}^\text{Haar}}{(\tilde{F}_t^\text{Haar})^2}-1\right].
\end{align}
We convert these again into confidence bounds using Cantelli's inequality and plot the results in Fig.~\ref{fig:frame_data}.

\subsection{Experimentally applying random unitaries}
\begin{figure*}[!t]\centering 
    \centering
    \includegraphics[width=1.0\columnwidth]{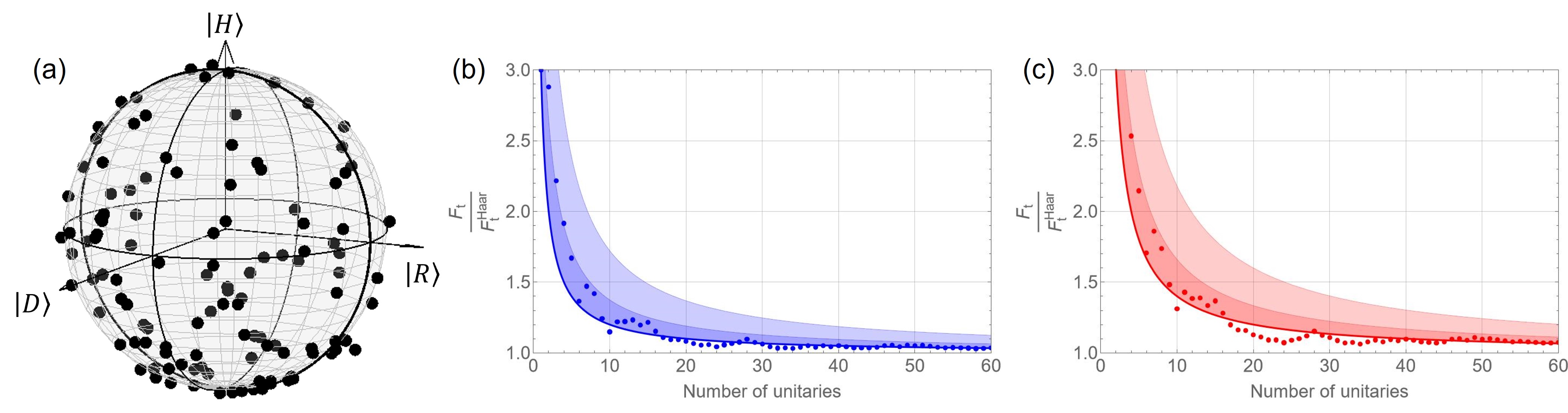}
    \caption{(a) Bloch sphere depicting the Stokes vectors resulting from 60 different unitary rotations applied with the polarization scrambler.
    (b, c) Experimentally measured spherical frame potential (points) along with the expected value (lines) given by Eq.~(\ref{eq:expGt}) for Haar-randomly drawn sets of unitaries as a function of the set size for t = 2 (b) and t = 4 (c). The colored areas above the curve denote the $1\sigma$ and $2\sigma$ confidence regions. As our main concern is a distribution of unitaries that yields larger frame potentials as expected, we give one-sided confidence regions.}
    \label{fig:frame_data}
\end{figure*}

We implement random local unitary rotations in the form of random polarization state rotations using polarization scramblers. 
One can naively expect that a polarization scrambler which can completely cover the Bloch sphere in a non-periodic fashion would indeed act as a random unitary rotation in polarization.
As such, we used a polarized light source and a standard polarimeter to inspect the polarization state output by the polarizaton scrambler after scrambling the polarization state 60 times.
The resulting polarization states are shown in Fig. \ref{fig:frame_data}(a).
Each polarization state measurement (denoted by each point on the sphere) was measured after scrambling for 10 seconds.

After measuring the polarization state output by the scrambler, we calculated the spherical frame potential. 
Figs. \ref{fig:frame_data}(b, c) show the resulting value of $F_{t}/F_{t}^{\text{Haar}}$ for a set of 60 randomly scrambled polarization states along with the $1\sigma$ and $2\sigma$ confidence regions described earlier.
The values of $F_{t}/F_{t}^{\text{Haar}}$ clearly fall within the confidence bounds, confirming that, with high probability, the polarization scrambler produces nearly-Haar random unitary rotations.

\section{Construction of unbiased estimators}
\label{app:unbiased_estimators}
The experimental data allows us to estimate the expectation values of the chosen observables. However, we want to average certain powers of this quantity to form the moments. This requires the usage of unbiased estimators.
\subsection{General results}
Suppose that we perform randomized measurements on two particles using the local observable $A\otimes A$ with
\begin{align}
    A = \text{diag}(k_1,k_2,\cdots,k_n).
\end{align}
In a single trial, we obtain one measurement outcome 
$X_i$, which is an element of the set $\{k_jk_l : 1\leq j \leq l \leq n\}$,
with corresponding outcome probability $p_i$ for $N$ independent trials.
Then one can write  
\begin{align}
    E =\mathrm{tr} \left[(U_A \otimes U_B) \rho_{AB} (U_A^{\dagger} \otimes U_B^{\dagger})(A \otimes A)\right] = \sum_i p_i X_i.
\end{align}

First, the unbiased estimator of $E$ is given by
\begin{align}
\widetilde{E} = \sum_i X_i \widetilde{p_i},
\end{align}
where $\widetilde{p_i} = {N_i}/{N}$ and $\mathbb{E}[\widetilde{p_i}]=p_i$.
Here, $N_i$ are the number of each event observed over the $N$ trials with $\sum_i N_i = N$.
Notice that $N_i$ are random variables following a multinomial distribution with parameters $(p_i, N)$.
One can immediately check that $\widetilde{E}$ is the unbiased estimator of $E$, that is, $\mathbb{E} \big[\widetilde{E}\big]=E$, by recalling the assumption that the $N$ trials are independent and $\mathbb{E} [N_i] = Np_i$.

Next, let us create an unbiased estimator $\widetilde{E^2}$ such that
$\mathbb{E}\big[\widetilde{E^2}\big] =E^2$.
It should be noted that $\widetilde{E^2}$ is not given by $\big(\widetilde{E}\big)^2$.
In fact, it can be written as
\begin{align}
    \widetilde{E^2} = \sum_i X_i^2 \widetilde{p_i^2}
    +2\sum_{i < j} X_i X_j \widetilde{p_ip_j},
    \label{eq:uE2}
\end{align}
where the unbiased estimators $\widetilde{p_i^2}$ and $\widetilde{p_ip_j}$ such that
$\mathbb{E}\big[\widetilde{p_i^2}\big]=p_i^2$ and 
$\mathbb{E}\big[\widetilde{p_ip_j}\big]=p_ip_j$
are given by
\begin{align}
    \widetilde{p_i^2} = \frac{N(\widetilde{p_i})^2-\widetilde{p_i}}{N-1},\ \ \
    \widetilde{p_ip_j}=\frac{N}{N-1}\widetilde{p_i} \widetilde{p_j}.
\end{align}
This can be straightforwardly shown using results from Ref.~\cite{newcomer2008computation}.

Similarly, we can create the unbiased estimators
$\widetilde{E^3}$ and $\widetilde{E^4}$ such that
$\mathbb{E}\big[\widetilde{E^3}\big] =E^3$ and
$\mathbb{E}\big[\widetilde{E^4}\big] =E^4$.
Using again Ref.~\cite{newcomer2008computation}, a straight-forward calculation leads to the expressions
\begin{align}
    \widetilde{E^3}
    &= \sum_i X_i^3 \widetilde{p_i^3}
    + 3\sum_{i<j}
    \left(
    X_i^2 X_j \widetilde{p_i^2 p_j}
    + X_i X_j^2 \widetilde{p_i p_j^2}
    \right)
    +6\sum_{i<j<k}
    \left(
    X_i X_j X_k
    \widetilde{p_i p_j p_k}
    \right),\\
    \widetilde{E^4}
    &= \sum_i X_i^4 \widetilde{p_i^4}
    + 4\sum_{i<j}
    \left(
    X_i^3 X_j \widetilde{p_i^3 p_j}
    + X_i X_j^3 \widetilde{p_i p_j^3}
    \right)
    +6\sum_{i<j}
    \left(
    X_i^2 X_j^2 \widetilde{p_i^2 p_j^2}
    \right) \nonumber\\
    \quad &+12\sum_{i<j<k}
    \left(
    X_i^2 X_j X_k \widetilde{p_i^2 p_j p_k}
    +X_i X_j^2 X_k \widetilde{p_i p_j^2 p_k}
    +X_i X_j X_k^2 \widetilde{p_i p_j p_k^2}
    \right)
    +24\sum_{i<j<k<l}
    \left(
    X_i X_j X_k X_l
    \widetilde{p_i p_j p_k p_l}
    \right),
\end{align}
where
the unbiased estimators
$\widetilde{q} \in \Big\{
\widetilde{p_i^3},
\widetilde{p_i p_j^2},
\widetilde{p_ip_jp_k},
\widetilde{p_i^4},
\widetilde{p_i^3 p_j},
\widetilde{p_i p_j^3},
\widetilde{p_i^2 p_j^2},
\widetilde{p_i^2 p_j p_k},
\widetilde{p_i p_j^2 p_k},
\widetilde{p_i p_j p_k^2},
\widetilde{{p_i p_j p_k p_l}}
\Big\}$
such that
$\mathbb{E}\big[\widetilde{q}\big]=q$
are given by
\begin{align}
    \widetilde{p_i^3}
    &= \frac{N^2\widetilde{p_i}^3-3N\widetilde{p_i}^2-2\widetilde{p_i}}{(N-1)(N-2)},\\
    \widetilde{p_i p_j^2}
    &= \frac{N^2\widetilde{p_i}^2\widetilde{p_j}-N\widetilde{p_i}\widetilde{p_j}}{(N-1)(N-2)},\\
    \widetilde{p_ip_jp_k}
    &= \frac{N^2\widetilde{p_i}\widetilde{p_j}\widetilde{p_k}}{(N-1)(N-2)},\\
    \widetilde{p_i^4}
    &= \frac{N^3\widetilde{p_i}^4-6N^2\widetilde{p_i}^3+11N\widetilde{p_i}^2-6\widetilde{p_i}}{(N-1)(N-2)(N-3)},\\
    \widetilde{p_i^3 p_j}
    &= \frac{N^3\widetilde{p_i}^3\widetilde{p_j}-3N^2\widetilde{p_i}^2\widetilde{p_j}+2N\widetilde{p_i}\widetilde{p_j}}{(N-1)(N-2)(N-3)},\\
    \widetilde{p_i p_j^3}
    &=\frac{N^3\widetilde{p_i}\widetilde{p_j}^3-3N^2\widetilde{p_i}\widetilde{p_j}^2+2N\widetilde{p_i}\widetilde{p_j}}{(N-1)(N-2)(N-3)},\\
    \widetilde{p_i^2 p_j^2}
    &=\frac{N^3\widetilde{p_i}^2\widetilde{p_j}^2-N^2(\widetilde{p_i}^2\widetilde{p_j}+\widetilde{p_i}\widetilde{p_j}^2)
    +\widetilde{p_i}\widetilde{p_j}}{(N-1)(N-2)(N-3)},\\
    \widetilde{p_i^2 p_j p_k}
    &=\frac{N^3\widetilde{p_i}^2\widetilde{p_j}\widetilde{p_k}-N^2\widetilde{p_i}\widetilde{p_j}\widetilde{p_k}}{(N-1)(N-2)(N-3)},\\
    \widetilde{p_i p_j^2 p_k}
    &=\frac{N^3\widetilde{p_i}\widetilde{p_j}^2\widetilde{p_k}-N^2\widetilde{p_i}\widetilde{p_j}\widetilde{p_k}}{(N-1)(N-2)(N-3)},\\
    \widetilde{p_i p_j p_k^2}
    &=\frac{N^3\widetilde{p_i}\widetilde{p_j}\widetilde{p_k}^2-N^2\widetilde{p_i}\widetilde{p_j}\widetilde{p_k}}{(N-1)(N-2)(N-3)},\\
    \widetilde{p_i p_j p_k p_l}
    &=\frac{N^3\widetilde{p_i}\widetilde{p_j}\widetilde{p_k}\widetilde{p_l}}{(N-1)(N-2)(N-3)}.
\end{align}

\subsection{Specific unbiased estimators for our experiment}

In our experimental investigation, we estimate the values of $I_{1}$, $I_{2}$, and $I_{3}$ from finite measurement results over observables $\mathcal{M}_{z}=\sigma_{3}\otimes\sigma_{3}$ and $\mathcal{M}_{det}=\sum_{i=1}^{3}\sigma_{i}\otimes\sigma_{i}$.
We assume in our experiment that we collect $N$ measurement results for each of $M$ pairwise random local unitary rotations. 
Each local projective measurement has two possible outcomes $X_{i}\in\{-1,1\}$; hence, we have four possible pairwise measurement outcomes.
For simplicity we will use the notation
\begin{align}
    E^{t}\left(\mathcal{M}_{i}\right) =\{\mathrm{tr} \left[(U_A \otimes U_B) \rho_{AB} (U_A^{\dagger} \otimes U_B^{\dagger})\mathcal{M}_{i}\right]\}^{t}
\end{align}

From the above results we can create an unbiased estimator for Eq.~(\ref{eq:R2ZZ}) in the main text  as
\begin{align}
\widetilde{I_{2}}=\widetilde{\RR{2}{Z\otimes Z}}=\frac{1}{M}\sum^{M}_{m=1}\left[\widetilde{E^{2}_{m}\left(\mathcal{M}_{z}\right)}\right]
\label{eq:app_I2_estimator}
\end{align}
where $m$ indicates the $m$-th set of local unitary rotations used to evaluate $E^{t}$.
Similarly, $\widetilde{I}_{3}$ can be accessed through a combination of $\widetilde{I_{2}}$ and the unbiased estimator of Eq.~(\ref{eq:R4ZZ})  in the main text which has the form
\begin{align}
\widetilde{\RR{4}{Z\otimes Z}}=\frac{1}{M}\sum^{M}_{m=1}\left[\widetilde{E^{4}_{m}(\mathcal{M}_{z})}\right].
\label{eq:app_R4_estimator}
\end{align}
Finally, we estimate $I_{1}$ directly through the unbiased estimator of Eq.~(\ref{eq:R3Adet})  in the main text which has the form
\begin{align}
  \widetilde{I_{1}}=\widetilde{\mathcal{R}_{\mathcal{M}_{det}}^{(3)}}=\frac{1}{M}\sum_{m=1}^{M}\left[\widetilde{E_{m}^{3}(\mathcal{M}_{det})}\right].
  \label{eq:app_I1_estimator}
\end{align}

\section{Statistical bounds}\label{app:xiaodong}

In order to derive confidence levels for the quantities that we derive, we apply Hoeffding's inequality, stating that for a set of $n$ statistically independent random variables $\{X_1, \ldots, X_n\}$, where each $X_i \in [a_i, b_i]$, the probability of observing their sum deviating more than $\delta$ from the mean value is bounded by \cite{hoeffding1994probability}
\begin{align}
    P\left(\vert \sum_i X_i - \mathbb{E}[\sum_i X_i] \vert \geq \delta\right) \leq \exp\left(-\frac{2\delta^2}{\sum_i (b_i - a_i)^2}\right).
\end{align}
We will use this inequality for independent, subsequent runs of the same inequality, i.e., $a_i \equiv a$, $b_i \equiv b$ for all $i$, and are interested in the mean, which demands a rescaling of $X_i \rightarrow X_i / n$.  By fixing the right-hand side to a desired error probability $1 - \gamma$, where $\gamma$ denotes the confidence, we obtain the following upper bound for the deviation $\delta$:
\begin{align}
     \delta = \frac{b-a}{\sqrt{2n}} \sqrt{\ln(\frac{2}{1-\gamma})}.
\end{align}
This allows us, for any given estimator of an experimental quantity which yields numbers in the range $[a,b]$, to derive the maximal deviation of the true mean value given a certain confidence level of $\gamma$. We will use this to bound the LU invariants $\trace(TT^\text{T})$, $\trace(TT^\text{T}TT^\text{T})$ and $\det(T)$, needed to determine the roots of the characteristic polynomial in Eq.~(\ref{eq:charpol})  in the main text.

What is left is to translate the confidence region of these coefficients into confidence regions of the roots. For that, we first give a naive bound on the confidence level that all of the measured quantities lie within their individual confidence levels.

\begin{figure}
    \centering
    \includegraphics[width=0.6\columnwidth]{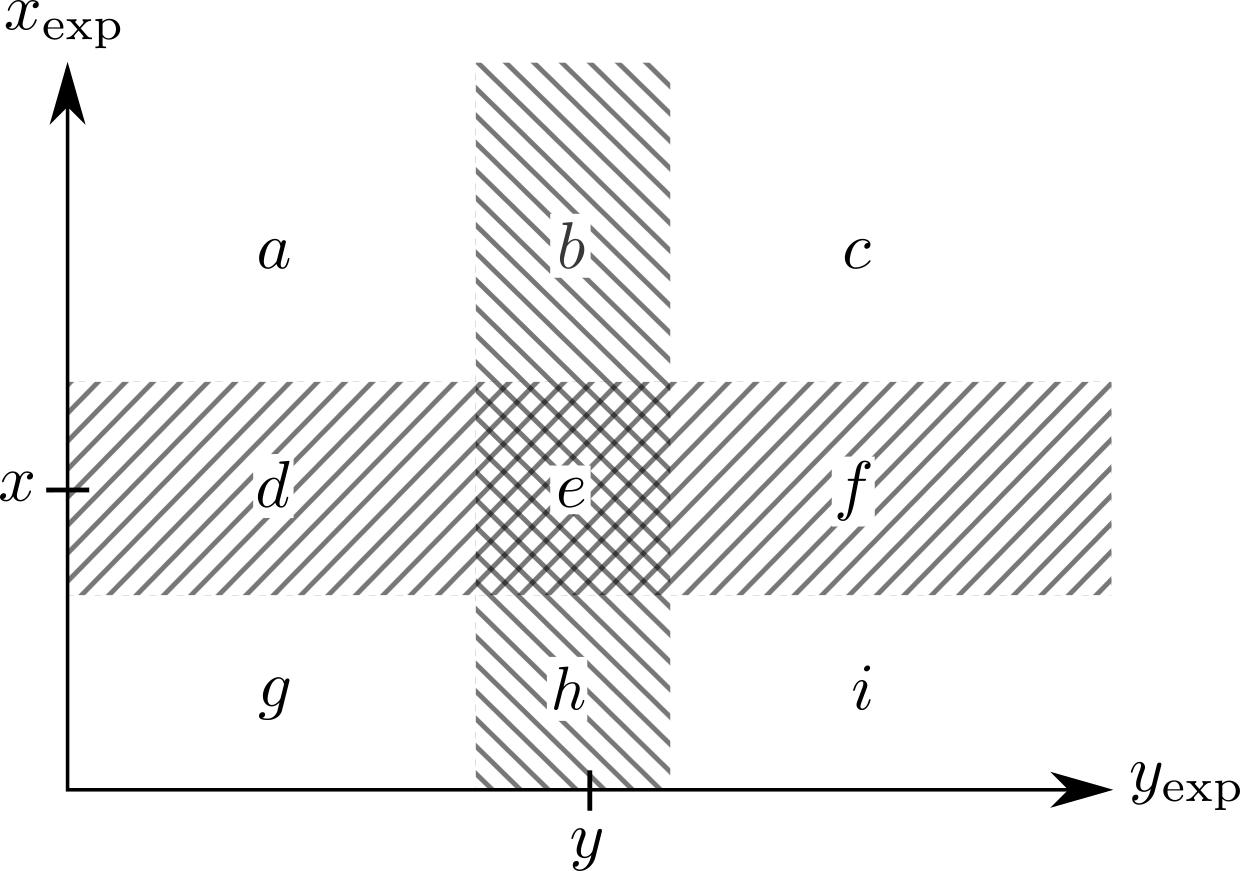}
    \caption{Illustration of the estimate in the text for two random variables: The probability to find experimental values inside of region $e$ is lower-bounded by the expression in Eq.~(\ref{eq:pbound}).  }
    \label{fig:confidences}
\end{figure}

To that end, consider first two random variables $x$ and $y$, the experimental values of which being with confidence level $\gamma$ in the regions $[x_0, x_1]$ and $[y_0,y_1]$, respectively. This is depicted in Fig.~\ref{fig:confidences}, where the confidence intervals of the two variables split the graph into $9$ different regions, where the symbols $a,b,\ldots,i$ denote the probability to find a pair of measurement results in the corresponding region. We have $d+e+f = b+e+h = \gamma$ and $a+b+c+g+h+i = a+c+d+f+g+i = 1-\gamma$. The probability to find an experimental value outside of at least one of the confidence intervals is given by $a+b+c+d+f+g+h+i = 2(1-\gamma) - (a+c+g+i) \leq 2(1-\gamma)$. Therefore, the probability to have measurement results within both confidence bands is at least $1-2(1-\gamma)$. The same argument can be applied to more than two variables, $x^{(1)}, \ldots, x^{(n)}$, where each variable $x^{(i)}$ lies in the interval $[x^{(i)}_0,x^{(i)}_1]$ with confidence $\gamma$, yielding
\begin{align}\label{eq:pbound}
    p(\forall i\,:\,x^{(i)}_0 \leq  x^{(i)} \leq x^{(i)}_1) \geq 1-n(1-\gamma).
\end{align}

Finally, we determine the expression in Eq.~(\ref{eq:chshviolation})  in the main text for each choice of coefficients in their corresponding confidence regions to obtain a range of violations, in which the true violation lies with confidence level $1-3(1-\gamma)$.

\section{Data analysis}\label{app:dataanalysis}

\subsection{Calculating CHSH violations from the measured data}

In order to certify the achieved violation of the CHSH inequality using Eq.~(\ref{eq:chshviolation})  in the main text  we first have to determine appropriate confidence intervals for the three LU invariants $\trace(TT^T)$, $\det(T)$ and $\trace(TT^TTT^T)$, respectively. To do so, we follow two approaches, the first of which assumes that the coarse-graining of all $25 \times 200$ data points in $25$ groups is enough to justify that the resulting data be normally distributed. Calculating the mean values and standard deviations based on this assumption yields
\begin{align}
    \det(T) &= -0.62\pm 0.15,\\
    \trace(TT^T) &= \phantom{-}2.41 \pm 0.15,\\
    \trace(TT^TTT^T) &= \phantom{-}2.21 \pm 0.21,
\end{align}
with $3\sigma$, i.e., 99.73\% confidence levels.

Alternatively, we can drop the assumption of normally distributed data and use the Hoeffding inequality to determine appropriate error bounds \cite{hoeffding1994probability}. For instance, the  sector length $\trace(TT^T)$ can be directly expressed as a sample mean $\sum_{i=1}^M X_i/M$ of the squared correlation function $X_i =\langle {U_A}^\dagger Z U_A \otimes {U_B}^\dagger Z U_B\rangle^2= 9 (p^{(i)}_{00} - p^{(i)}_{01} - p^{(i)}_{10} + p^{(i)}_{11})^2$, with $X_i\in[0, 9]$. Applying the Hoeffding inequality to this case allows us to assign, with confidence $\gamma$, the following two-sided error bound:
\begin{align}
    \delta = \frac{9}{\sqrt{2M}} \sqrt{\ln(\frac{2}{1-\gamma})},
\end{align}
which for $M=25\times 200 = 5000$ and $\gamma = 0.9973$ ($3\sigma$) leads to
\begin{align}
    \trace(TT^T) = 2.41 \pm 0.24.
\end{align}
Using the Hoeffding inequality for the determinant, we obtain
\begin{align}
    \det(T) = -0.62\pm 1.09,
\end{align}
which is significantly worse compared to the Gaussian estimate.
Lastly, for $\trace(TT^TTT^T)$, we cannot apply the same arithmetic, as this quantity does not originate directly from a sample average over the runs but instead also involves the square of the respective sector length  
$\trace(TT^T)^2$. As a workaround, we exploit the  insight that the range of physically allowed values of the quantity $\trace(TT^TTT^T)$ is constrained by the value of $\trace(TT^T)$. Thus, we can derive a region of compatible values of $\trace(TT^TTT^T)$ from the confidence region of $\trace(TT^T)$, i.e.   $\trace(TT^T) = 2.41 \pm 0.24$. Following this procedure, we obtain
\begin{align}
    \trace(TT^TTT^T) = 2.00 \pm 0.42.
\end{align}

We now have two sets of results including $3\sigma$ confidence levels for the three invariants under consideration. We can thus proceed to calculate the roots of the characteristic polynomial~Eq.~(\ref{eq:charpol})  in the main text and, respectively, the achievable violation of the CHSH inequality. In order to determine the best permissible value of the latter, we scan the whole range of allowed values of the three invariants and  calculate the corresponding roots and CHSH violation for each of them. Finally, we use the largest violation which is still  compatible with the observed data. The confidence of this violation is then given by $1-3(1-\gamma) = 0.991 \approx 2.6\sigma$, as detailed in Appendix~\ref{app:xiaodong}. 

Following the above procedure, we obtain the following CHSH violations:
\begin{align}
    \text{CHSH}_\text{Gauss} &\geq 0.46, \\
    \text{CHSH}_\text{Hoeff} &\geq 0.40.
\end{align}

Note that the maximal observable value is given by $2\sqrt{2} - 2 \approx 0.83$.

Similarly, by requiring a higher confidence level of $5\sigma$ for the invariants, or equivalently $\gamma = 0.9999994$, we obtain
\begin{align}
    \text{CHSH}_\text{Gauss} \geq 0.42, \\
    \text{CHSH}_\text{Hoeff} \geq 0.34,
\end{align}
with confidence $1-3(1-\gamma)=0.999998\approx 4.7\sigma$. 
Using either method, our results clearly show that the randomized measurement protocol successfully determines that the state output by our EPS has the potential to violate a CHSH inequality.

\subsection{Bounding the maximal teleportation fidelity from below}
The same error analysis can be used to obtain the lower bound $F_{\text{max}}^{U}$ 
in Eq.~(\ref{eq:maxfidelity})  in the main text from the confidence intervals of the LU invariants.
For a confidence level of $3\sigma$ of the LU invariants, we obtain a lower bound of at least:
\begin{align}
    \left(F_{\text{max}}^{U}\right)_{\text{Gauss}}&=0.88,\\
	\left(F_{\text{max}}^{U}\right)_{\text{Hoeff}}&=0.85,
\end{align}
with a confidence confidence level of $1-3(1-\gamma) = 0.991 \approx 2.6\sigma$.

Similarly, we obtain the lower bounds
\begin{align}
    \left(F_{\text{max}}^{U}\right)_{\text{Gauss}}&=0.86,\\
	\left(F_{\text{max}}^{U}\right)_{\text{Hoeff}}&=0.60,
\end{align}
for a confidence level of $5\sigma$ of the LU invariants.
The confidence level of the bounds is $1-3(1-\gamma)=0.999998 \approx 4.7\sigma$.

\subsection{Calculating the expected values from quantum state tomography}

Here, we provide a more-detailed description of how the expected values for each invariant, the CHSH violation, and the teleportation fidelity are calculated from quantum state tomography. For every randomized measurement performed (200 unitaries x 25 runs), a corresponding quantum state tomography was performed for benchmarking the randomized measurement results. The expected values of $I_{1}$, $I_{2}$, and $I_{3}$ were then calculated from the density matrices resulting from quantum state tomography via Eq.~(A2), and the results were averaged over every density matrix. The green bands in Fig.~1(b-d) show the average value plus or minus the standard deviation for each invariant: $I_{1} = -0.71\pm0.12$, $I_{2} = 2.41\pm0.34$, and $I_{3} = 1.95\pm0.34$.

Next, the concurrence of each density matrix was calculated and the maximal CHSH violation, $\text{CHSH}_{\text{QST}} = 2-2\sqrt{1+C^{2}}$, was calculated from the expression $S=2\sqrt{1+C^{2}}$ \cite{horodecki1995violating}. This analytical relationship between $S$ and $C$ holds for rank-2 Bell diagonal states, and we have demonstrated in our previous work \cite{jones2018tuning, kirby2018effect, jones2020exploring} that the state output by our system can indeed be approximated by such a state. After averaging over all density matrices, we determined a maximal CHSH violation of $\text{CHSH}_{\text{QST}} \leq 2.60\pm0.11$. Of course, due to noise and the imperfect nature of any experimental system, the actual state output by our source is not exactly a Bell diagonal state and the expected CHSH violation is less than the upper-bound represented by this expression. Therefore, the CHSH violation of $\text{CHSH}\geq 0.46$ determined by randomized measurements is in agreement with the maximal violation determined from tomography and clearly shows that the randomized measurement protocol successfully verifies the nonlocal nature of states output by our EPS. Finally, the fidelity of the states determined by tomography were averaged, and we determined that $F_{\text{QST}} = 0.90\pm0.08$. We then calculated the teleportation fidelity $f_{\text{QST}} = 0.93\pm0.05$ using Eq.~(13).

Since the experiment was performed for several hours per day over several days, variation in the performance of the entangled photon source resulted in a rather large standard deviation of the expected values for each invariant. The variation in each of the values is consistent with the typical performance of our system, which has previously been thoroughly characterized \cite{jones2018tuning, kirby2018effect, jones2020exploring}. Furthermore, we emphasize that the variation in the expected values calculated from tomography is comparable to the variation in the values determined by randomized measurements. This suggests that much of the uncertainty in the values determined by randomized measurements is due to variation/drift in the performance of the entangled photon source. Therefore, we claim that the randomized measurement scheme offers sufficient measurement accuracy with the significant advantage that it can be used in practical cases where it would be difficult or impossible to align all of the measurement bases required for full quantum state tomography.

\end{document}